\documentclass[a4paper]{article}
\usepackage{lineno}
\usepackage[utf8]{inputenc}
\usepackage{indentfirst}
\usepackage{xcolor}
\usepackage[T1]{fontenc}
\usepackage{subcaption}
\usepackage{caption}
\usepackage{afterpage}
\usepackage{float}
\usepackage[a4paper,top=3cm,bottom=2cm,left=3cm,right=3cm,marginparwidth=1.75cm]{geometry}
\usepackage{amsmath}
\usepackage{graphicx}
\usepackage[colorinlistoftodos]{todonotes}
\usepackage[colorlinks=true, allcolors=blue]{hyperref}
 
\title{Learning Surrogate Rainfall-driven Inundation Models with Few Data}
\author{
  \small{  Marzieh Alireza Mirhoseini\thanks{Email: malireza@mit.edu}}\\
    Department of Earth Atmospheric and Planetary  Sciences \\
      Massachusetts Institute of Technology, Cambridge, MA 02139, USA, 
}
\usepackage{todonotes}

\begin{document}
% \linenumbers

\maketitle
\renewcommand{\figurename}{Fig.}

\begin{abstract}

Flood hazard assessment demands fast and accurate predictions. Hydrodynamic models are detailed but computationally intensive, making them impractical for quantifying uncertainty or identifying extremes. In contrast, machine learning surrogates can be rapid, but training on scarce simulated or observed extreme data can also be ineffective.
This work demonstrates that we can develop an effective surrogate model for flood hazard prediction by initializing deep learning (ResNET-18) with ensemble-approximated Conditional Gaussian Processes (EnsCGP) and finalizing it with a bias correction. The proposed methodology couples EnsCGP with a ResNet-18 architecture to estimate flood depth and uses an ensemble optimal estimation for bias correction.
The surrogate model was trained and evaluated using rainfall data from Daymet and hydrodynamic simulations from LISFLOOD-FP, spanning the period between 1981 and 2019. The training involved using data up to a certain year and testing on the subsequent year, iteratively progressing through the dataset. This process required approximately 100 training iterations and extensive data. Inundation depths are estimated rapidly at run-time ($\sim 0.006$ seconds per event). Results over multiple years in the current climate over Chicago demonstrate an average $R^2 > 0.96$, with median relative errors in flood depth estimates of about 1\%.
% 1 percentage is came from 0.01/0.3 ~ 1%
\end{abstract}

\section{Introduction}

Flood risk assessment is essential for understanding the potential impacts of flooding on people, infrastructure, and the environment. It shares tools and methods with flood forecasting, such as remote sensing, rainfall measurements, and hydrodynamic models~\cite{De_Roo_2000, Prakash2021Flood, Jiang2021Hydrodynamic}. However, unlike flood forecasting—which focuses on predicting specific flood events in real-time—flood risk assessment adopts a broader perspective by incorporating historical flood records and climate projections to determine flood hazards. It further evaluates long-term exposure and vulnerability, integrating diverse data sources over extended periods and across different regions to identify high-risk areas and vulnerable populations~\cite{Manfreda2019Digital, Pham2021FloodRisk}. By providing an understanding of the probability of future flooding~\cite{De_Roo_2000, VanDerKnijff2010LISFLOOD, Manfreda2019Digital, Bates2010LISFLOODFP}, flood risk assessments enable the development of long-term strategic planning and mitigation measures to address future vulnerabilities.

Although challenges exist across the hazard, exposure, and vulnerability components of the flood risk chain, the computational expense of hydrodynamic models used in hazard modeling is particularly significant, especially in large-scale or urban environments. Inundation models require vast amounts of high-resolution data and extensive processing time, making them less practical for large-scale risk assessments~\cite{LopezLopera2022Computational}. Robust risk quantification often necessitates numerous ensemble Monte Carlo simulations across various scenarios to adequately capture uncertainty and assess risk.

To address these limitations, there is growing interest in more efficient and adaptable approaches, such as surrogate models, which reduce computational demands while maintaining accuracy across diverse regions~\cite{Ni2023Data, Lee2023Surrogate}. In this paper, we focus on inundation modeling in urban environments—a critical element of the flood risk chain—to estimate water depths from rainfall while incorporating various parameters. The LISFLOOD-FP~\cite{Bates2010LISFLOODFP} model series serves as an example of a modern flood inundation model used for this purpose.

Comprehensive flood risk analyses often require running thousands of simulations for stochastically sampled synthetic timelines of extreme events. Therefore, having a fast and accurate inundation model is highly desirable. Increased computational speed would enhance not only risk assessments but also uncertainty quantification and the identification of extreme scenarios. However, current models are computationally intensive; for instance, a single LISFLOOD-FP simulation over the larger Chicago area (see Figure~\ref{fig:area}
a) takes approximately 20 minutes on an NVIDIA RTX 2080 Ti GPU. While faster alternatives exist, achieving significant improvements in speed without compromising the fidelity of numerical simulations remains a fundamental challenge.

Machine Learning could offer an approach, mainly if relatively few detailed hydrodynamic simulations can train a surrogate. If needed, this surrogate could be further adapted with observed data to overcome biases in numerical simulations. It would also be beneficial if the retraining from location to location could be minimal. This paper develops an approach to the first of these three directions. It proposes a surrogate model that primes learning using a relatively shallow machine (ResNet-18) ~\cite{he2016deep} with ensemble-approximated Conditional Gaussian Processes (EnsCGP) \cite{ravela2007fast,ravela2010realtime,ravela2020informative}. The statistical-neural architecture follows earlier work using a similar priming mechanism for 
 downscaling~\cite{saha2024statistical}, where it was observed that the statistical priming ameliorates data needs during training. 

 The model is trained on Daymet rainfall data and LISFLOOD-FP flood simulations and applies  EnsCGP for initial predictions. The model also incorporates a ResNet-18 architecture, a deep residual network~\cite{he2016deep}. The initial guess is computed using a reduced rank EnsCGP (Section \ref{sec:GP}) in approximately $0.01$ seconds per event with the same computing setup. The ResNet refines the initial prediction made by EnsCGP, delivering a final estimate within 0.001 seconds for the downtown Chicago region (Fig.~\ref{fig:area}b). Bias correction through the Empirical Cumulative Distribution Function (ECDF) \cite{THORNTON1997214} is finally applied. The overall workflow is illustrated in Fig.~\ref{fig:flowchart}.
 
This methodology provides an accurate and computationally efficient for quantifying flood risk, addressing some limitations in existing approaches.   As the results on emulating urban floods in Chicago demonstrate, the model consistently delivers accurate flood depth estimates with minimal computational demand \cite{adriano2023combining, Elbaida2024, ALIZADEHKHARAZI2021101628}.

The remainder of the paper is organized as follows: Section \ref{sec:2} details the study area and data sources, Section \ref{sec:3} outlines the development and integration of the surrogate model, Section \ref{sec:conclusion} evaluates model performance, and the conclusion highlights key findings and future research directions.

%\subsection{Traditional and Pure Data-Driven Approaches}
\section{Related Work}
\label{sec:rw}
%Data-driven Models
Data-driven models utilize historical flood data and observations to predict flood levels, offering a faster alternative to traditional hydrodynamic models. Chang et al. (2022) \cite{CHANG2022128086} and Lee et al. (2023) \cite{Lee2023Surrogate} demonstrated that these models can quickly predict flood levels using techniques like Principal Component Analysis (PCA) and statistical analysis. While effective in many scenarios, these models rely heavily on the quality and quantity of available data, which can affect their performance in data-scarce regions.

%Machine Learning Models
Machine Learning (ML) techniques, such as Random Forest (RF) and K-Nearest Neighbors (KNN), have gained popularity in flood hazard assessment due to their ability to handle large datasets and capture complex relationships among factors like rainfall, topography, and flood inundation. Pham et al. (2021) \cite{PHAM2021106899} and Hou et al. (2021) \cite{Hou2021Urban} illustrated the versatility of ML models in flood mapping and forecasting, enabling rapid classification of flooded areas with high accuracy. The robustness of RF and KNN in handling noisy data and high-dimensional inputs was demonstrated by Zahura and Goodall (2022) \cite{ZAHURA2022101087}, Bayat and Tavakkoli (2022) \cite{bayat2022application}, and Luu et al. (2021) \cite{Luu2021Flood}, and Salas et al. (2023) \cite{SALAS2023118862}.

Recent advancements have improved the computational efficiency and accuracy of ML models in flood prediction. Zeng et al. (2023) \cite{zeng2023global} enhanced ML models through multimodal approaches, while Karim et al. (2023) \cite{karim2023review} provided a comprehensive review of automated flood prediction models. Techniques such as the Light Gradient Boosting Machine (LightGBM), applied by Xu et al. (2023) \cite{xu2023rapid}, and ensemble methods used by Rahman et al. (2023) \cite{rahman2023flood}, have further increased the ability of ML models to deliver rapid, reliable flood predictions.

%Deep Learning Models
Deep learning techniques, including Convolutional Neural Networks (CNNs) and Long Short-Term Memory (LSTM) networks, have significantly advanced flood hazard assessment by automating feature extraction and handling complex spatiotemporal relationships in data. Hu et al. (2019) \cite{HU2019911} and Chang et al. (2022) \cite{CHANG2022128086} demonstrated how neural networks can improve the temporal and spatial accuracy of flood forecasts, particularly in urban areas with limited data availability. Huang et al. (2023) \cite{Huang2023Flood} introduced the Flood-Precip GAN, a model capable of generating synthetic flood events, contributing to flood preparedness by augmenting datasets.

Physics-Informed Neural Networks (PINNs), originally introduced by Raissi et al. (2019) \cite{Raissi2019PINN}, represent a novel approach that embeds physical laws within deep learning models. This integration improves predictive performance and reduces computational demands. Donnelly et al. (2023) \cite{Donnelly2023PINN} explored the use of PINNs in flood modeling, highlighting how incorporating physical constraints enhances model performance and interpretability. However, deep learning models can sometimes lack transparency compared to physically based models, which is important for understanding hydrological processes driving flood events (Momoi et al., 2023 \cite{Momoi2023Physics}; Shao et al., 2023 \cite{shao2024advancing}).

%Hybrid Models
Hybrid models combine machine learning with traditional physics-based approaches, leveraging the strengths of both to achieve faster flood predictions without compromising physical accuracy. Hou et al. (2021) \cite{Hou2021Urban} improved urban flood forecasting by integrating RF and KNN algorithms with hydrodynamic models. Xu et al. (2023) \cite{Xu2023Hybrid} demonstrated the effectiveness of hybrid models by coupling ML techniques with physical models to enhance prediction accuracy.

The effectiveness of hybrid models is particularly evident in data-scarce regions, where combining ML with physically based models enhances flood hazard prediction (Van der Knijff et al., 2010 \cite{van2010lisflood}; Zanchetta et al., 2022 \cite{Zanchetta2022Improving}). Ensemble-based methods like MaxFloodCast, developed by Lee et al. (2023) \cite{Lee2023Surrogate}, and probabilistic forecasting using LSTM neural networks, as shown by Hop et al. (2023) \cite{Hop2023Probabilistic}, further demonstrate the potential of hybrid models to incorporate rainfall forecasts and hydrodynamic simulations for real-time flood prediction.

%Surrogate Models
Surrogate models, designed to emulate hydrodynamic simulations, have gained traction as computationally efficient alternatives to traditional flood modeling. Zhang et al. (2019) \cite{Zhang2019Surrogate}, Hosseiny et al. (2020) \cite{Hosseiny2020ANN}, and Bermúdez et al. (2018) \cite{Bermudez2018Computationally} highlighted the utility of surrogate models in reducing computational costs while maintaining accuracy in flood risk assessments. López-Lopera et al. (2022) \cite{LOPEZLOPERA2022108139} applied multi-output Gaussian Process models for coastal flood hazard prediction, offering both spatial and temporal efficiency.

Wei et al. (2023) \cite{Wei2023Rapid} developed an encoder-decoder LSTM model for rapid, high-resolution flood predictions. Adriano et al. (2023) \cite{adriano2023combining} created a hybrid surrogate model that integrates LSTM and 1D CNN architectures for coastal flood risk management, reflecting the growing trend of combining machine learning with hydrodynamic data to optimize flood predictions. 

\section{Study Area and Data}{\label{sec:2}}

\begin{figure}[h!]
    \centering
    \begin{tikzpicture}
        % Node for the first image
        \node (img1) at (0,0) {
            \begin{minipage}[b]{0.6\textwidth}
                \centering
                \includegraphics[width=\linewidth]{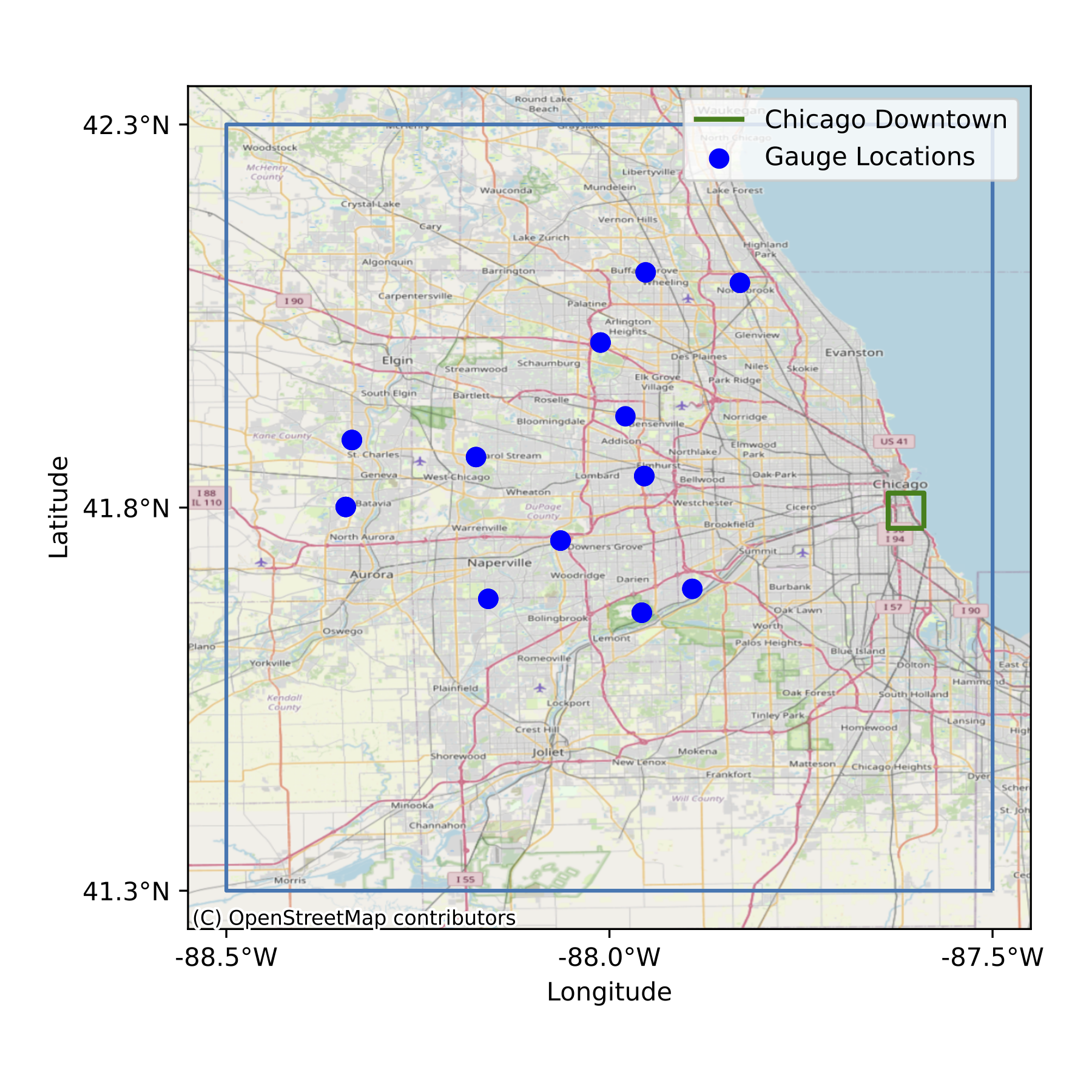}
              %run /net/flood/data/users/malireza/map_location_asli.py
                \caption*{(a)}
                \label{fig:map}
            \end{minipage}
        };
        
        % Node for the second image
        \node (img2) at (8,-1.2) {
            \begin{minipage}[b]{0.4\textwidth}
                \centering
                \includegraphics[width=\linewidth]{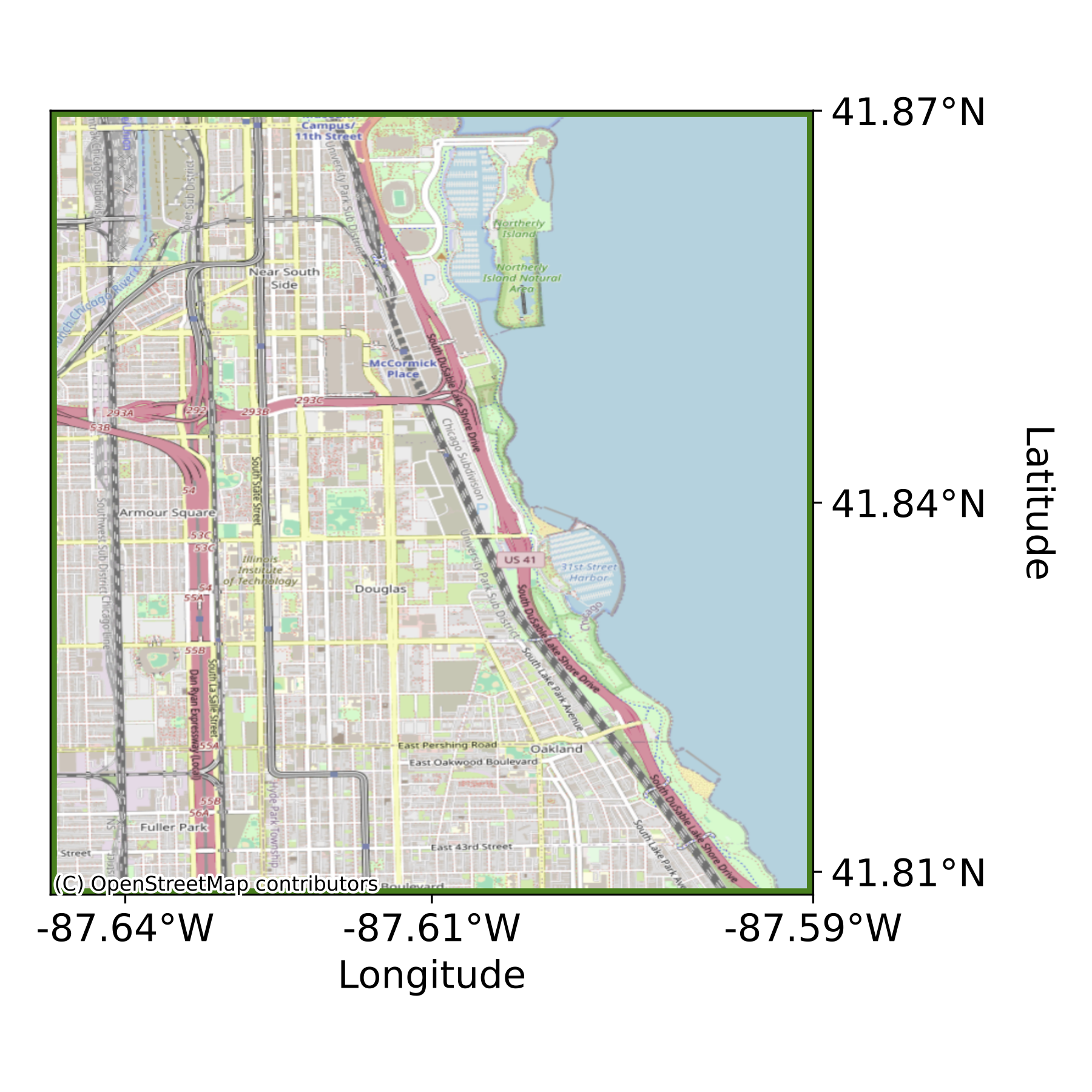}
                %run /net/flood/data/users/malireza/map_location_asli.py
                \caption*{(b)}
                \label{fig:crop}
            \end{minipage}
        };
        
        % Coordinates for the green box in the first image
        \coordinate (p1) at (3.14, 0.8); % Top left of green box
        \coordinate (p2) at (3.14, 0.55); % Bottom left of green box

        % Coordinates for the second image
        \coordinate (p5) at (5.1, 1.7); % Top center of second image
        \coordinate (p6) at (5.1, -2.9); % Bottom center of second image
        
        % Connecting lines
        \draw[thick] (p1) -- (p5);
        \draw[thick] (p2) -- (p6);
    \end{tikzpicture}
    \caption{Overview of the Chicago study area, illustrating the spatial distribution of gauge stations monitoring water depths across urban and suburban regions (a). The zoomed-in panel (b) offers a detailed view of the downtown Chicago area. Both maps are derived from OpenStreetMap data.}

    \label{fig:area}
\end{figure}

The study focuses on the Chicago metropolitan area in northeastern Illinois, USA, spanning latitudes $41.345^\circ \text{N}$ to $42.345^\circ \text{N}$ and longitudes $88.505^\circ \text{W}$ to $87.505^\circ \text{W}$. Figure~\ref{fig:area}a illustrates the entire region where numerical flood simulations were conducted using the LISFLOOD-FP hydrodynamic model. Figure~\ref{fig:area}b highlights a smaller cropped area within this region, where we have implemented a surrogate model (see Section~\ref{sec:3}).

Flood simulations for the region depicted in Fig.~\ref{fig:area}a were performed for the period 1981--2019 using the LISFLOOD-FP hydrodynamic model (version 8.1)~\cite{sharifian2022lisflood}, a GPU-accelerated model that solves the two-dimensional shallow water equations~\cite{tan1992shallow}. The simulations were driven by rainfall data from the Daymet dataset~\cite{thornton2014daymet}, which provides high-resolution coverage ($1\,\text{km} \times 1\,\text{km}$). The model employs a finite volume scheme and open boundary conditions to capture high-resolution flood dynamics in the study area. Terrain elevation data at 1-arc-second resolution from the Shuttle Radar Topography Mission (SRTM)~\cite{farr2007shuttle} and Manning's coefficients derived from the National Land Cover Database (NLCD)~\cite{homer2012national} were integrated into the model. The output comprises two-dimensional maps of daily maximum flood inundation.

\begin{figure}[htbp]
    \centering
    \begin{subfigure}[b]{0.55\linewidth}
        \includegraphics[width=\linewidth]{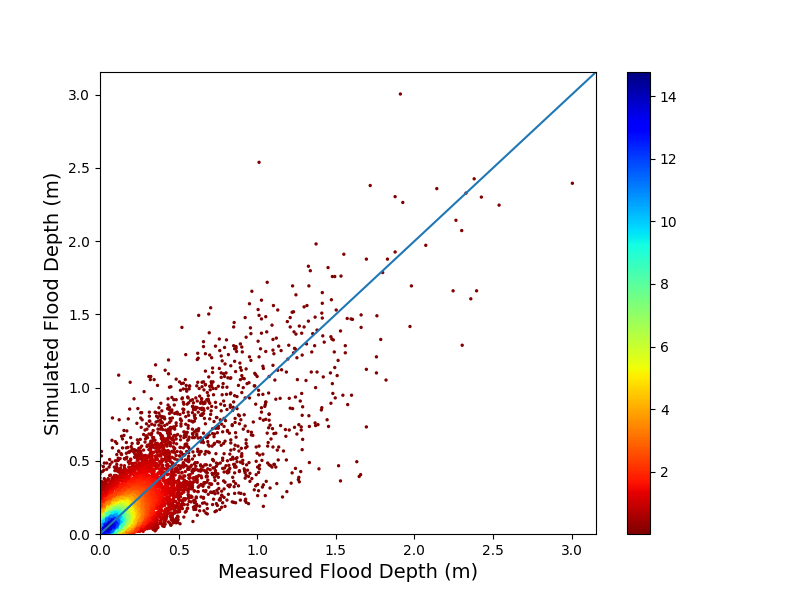}
         \caption{}
\end{subfigure}
        \begin{subfigure}[b]{0.42\linewidth}
\includegraphics[width=\linewidth]{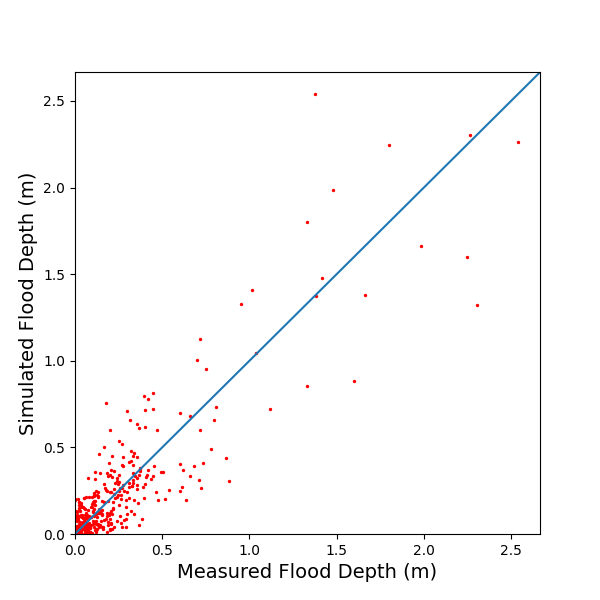}
\caption{}
\end{subfigure}
        \caption{Scatter plots comparing simulated and observed water depths for the gauge locations shown in Fig. \ref{fig:area} a. Plot (a) displays data from all years, while plot (b) focuses on the extreme flood year of 2013.}
        \label{fig:numerical_vs_gauge}
        %run code /net/flood/data/users/malireza/new_update_scatter_real_num.py
    \end{figure}

Model validation was performed by comparing simulated water depths with observed measurements from the National Water Information System (NWIS)~\cite{USGS_NWIS_2023}, with data contributions from the USGS and NOAA. The gauge locations used for validation are indicated in Fig.~\ref{fig:area}a. The hydrological data, specifically the daily maximum water depths, were retrieved and processed using the Hydrofunctions Python library~\cite{Roberge_Hydrofunctions_2023}. The scatter plots in Fig.~\ref{fig:numerical_vs_gauge} demonstrate strong agreement between observed and simulated water depths, confirming the reliability of the LISFLOOD-FP model for hydrodynamic simulation in this area.

In this study, we have implemented a surrogate model for the cropped region shown in Fig.~\ref{fig:area}b. The surrogate model aims to efficiently approximate the flood simulations produced by the LISFLOOD-FP model in this smaller area. As future work, we plan to extend the surrogate modeling approach to additional cropped regions within the study area, as indicated in Fig.~\ref{fig:area}b, to further evaluate its applicability and performance across different parts of the region.

Throughout this paper, ``flood simulation'' refers to the results obtained from the LISFLOOD-FP model.

% The surrogate model is trained against extreme rainfall events (90th percentile or higher) in Chicago, and corresponding flood responses. The rainfall events are extracted from Daymet dataset~\cite{thornton2014daymet}, a gridded daily rainfall product of 1 Km resolution. Flood responses to these extreme rainfall events were numerically simulated using LISFLOOD-FP 8.1~\cite{sharifian2022lisflood}, a GPU-accelerated hydrodynamic model that solves the two-dimensional shallow water equations~\cite{tan1992shallow}. The simulation employs a finite volume scheme and open boundary conditions to capture high-resolution flood dynamics in the study area. Terrain elevation data of 1 arc-second resolution from Shuttle Radar Topography Mission (SRTM)~\cite{farr2007shuttle} and Manning's coefficients derived from National Land Cover Database (NLCD) land cover data~\cite{homer2012national} were integrated into the model. The output of the hydrodynamic model are the two-dimensional maps of daily maximum flood inundation, which are calibrated and validated against measured water height at selected gauge locations. The gauge measurements are acquired from the National Water Information System (NWIS) \cite{USGS_NWIS_2023} using the  Hydrofunctions Python library \cite{Roberge_Hydrofunctions_2023}. The comparison between measured and simulated water depth is presented in Figure \ref{fig:numerical_vs_gauge}a, while \ref{fig:numerical_vs_gauge}b shows the same comparison for the year 2013, a specific year of interest.

%  \begin{figure}[htbp]
%     \centering
%     \includegraphics[width=0.6\linewidth]{guage_map.png}
%         \caption{Spatial distribution of gauge stations across the Chicago metropolitan region. The gauges are used to monitor flood depths and are strategically located to provide comprehensive coverage of urban and suburban areas.} % Add a caption for the second subFig.
%         \label{fig:gauge_map}
%     % run map_location_asli.py
%     \label{fig:gauge}
% \end{figure}

\usetikzlibrary{shapes.geometric, arrows, positioning, calc, fit}

\tikzstyle{block} = [rectangle, draw, fill=none, 
    minimum width=9em, minimum height=3em, text centered, align=center]
\tikzstyle{circle_block} = [circle, draw, fill=none, 
    minimum size=5.5em, text centered, align=center]  % Set align=center for multi-line text
\tikzstyle{bias_block} = [circle, draw, fill=none, 
    minimum size=5.5em, text centered, align=center]  % Different color for Bias Correction
\tikzstyle{line} = [draw, -latex']
\tikzstyle{surrogate_box} = [draw, dashed, inner sep=1em, rounded corners]

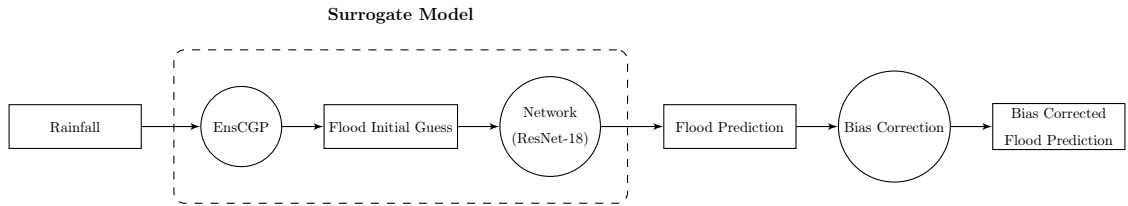
\begin{figure}[ht!]
    \centering
    \begin{tikzpicture}[node distance=1cm, scale=0.55, transform shape]
        % Place nodes
        \node [block] (block1) {Rainfall};
        % \node [block, below=of block1] (block2) {Numerical Results};
        \node [circle_block, right=of $(block1)$, xshift=2cm] (block3) {EnsCGP};
        \node [block, right=of block3] (block7) {Flood Initial Guess};
        \node [circle_block, right=of block7] (block4) {Network \\(ResNet-18)};
        \node [block, right=of block4, xshift=0.5cm] (block8) {Flood Prediction};  % New Flood Prediction block outside surrogate model box
        \node [bias_block, right=of block8] (block5) {Bias Correction};  % Bias Correction in different color
        \node [block, right=of block5] (block6) {Bias Corrected\\Flood Prediction};

        % Draw surrogate model box around Gaussian Process, Flood Initial Guess, and ResNet-18 nodes
        \node [surrogate_box, fit=(block3) (block4) (block7)] (surrogate_model) {};
        \node [above=of surrogate_model, yshift=-0.5cm] {\large \textbf{Surrogate Model}};

        % Draw edges
        % \path [line] (block1) -- (block2);
        \path [line] (block1) -- (block3);
        % \path [line] (block1) -- (block3);
        % \path [line] (block2) -- (block3);
        \path [line] (block3) -- (block7);
        \path [line] (block7) -- (block4);
        \path [line] (block4) -- (block8);  % Connect Network to Flood Prediction
        \path [line] (block8) -- (block5);  % Connect Flood Prediction to Bias Correction
        \path [line] (block5) -- (block6);
    \end{tikzpicture}

    \caption{ A schematic representation of the surrogate modeling approach for flood prediction. Daymet rainfall data from the test set is used as input to a Conditional Gaussian Process (Eq. \ref{Eq:GP}), which generates an initial guess of water depth. This initial guess is then refined by a ResNet-18 network, producing the flood prediction. Bias correction is subsequently applied to adjust the flood prediction, resulting in a final bias-corrected flood prediction.}
    \label{fig:flowchart}
\end{figure}

\section{Methodology}{\label{sec:3}}

In this study, we develop a surrogate model based on flood simulation for the region shown in Fig. \ref{fig:area}b, to estimate peak flood inundation more efficiently. The aim was to replicate the behavior of detailed physics-based flood simulations while minimizing computational time and effort.

The model training process began by using an ensemble-approximated Conditional Gaussian Process (EnsCGP) (Section \ref{sec:GP}) to generate initial predictions of water depths. These preliminary estimates were then refined through a ResNet-18 network (Section \ref{sec:resnet}), which was used to capture critical spatial patterns in the data and improve the accuracy of the flood predictions.

Once trained, the surrogate model was tested on new rainfall datasets. After generating predictions, we applied a bias correction step using the Empirical Cumulative Distribution Function (ECDF) (Section \ref{sec:bias}) to account for any systematic errors and to enhance the accuracy of the results further.

The complete workflow is shown in Fig.~\ref{fig:flowchart}, where it’s clear that the only input data used in this process is rainfall data.

\subsection{ Ensemble-Approximated Conditional Gaussian Process} \label{sec:GP}

In this study, we employ an ensemble-approximated Conditional Gaussian Process (EnsCGP) model to learn the relationship between rainfall data \(X=\{x_i\}_{i=1}^N\) and corresponding flood simulation \(Y=\{y_i\}_{i=1}^N\). The EnsCGP model utilizes the underlying covariance structure between \(X\) and \(Y\) to enable flood emulations \(y_q\) for new rainfall \(x_q\). The predictive model is given by:

\begin{equation}
    y_q - \bar{Y} = \mathrm{Cov}(Y,X) \mathrm{Cov}(X,X)^{-1} (x_q - \bar{X}),
\end{equation}
where $\bar{X}$ and $\bar{Y}$ are the mean vectors of the rainfall and numerical flood data from the training set. Instead of explicitly computing the covariance matrices, we directly use the relationship between the ensemble as:

\begin{equation}
    \mathrm{Cov}(X,X) = \frac{1}{N-1} (X - \bar{X}) (X - \bar{X})^T, \quad \mathrm{Cov}(Y,X) = \frac{1}{N-1} (Y - \bar{Y}) (X - \bar{X})^T.
\end{equation}

Given the high-dimensional nature of the data, Principal Component Analysis (PCA) was used to decompose the snapshot matrix:

\begin{equation}
    X - \bar{X} = U \Sigma V^T,
\end{equation}
where \(U \in {R}^{M \times M}\) and \(V \in {R}^{N \times N}\) are unitary matrices containing the orthonormal eigenvectors of \((X - \bar{X})(X - \bar{X})^T\) and \((X - \bar{X})^T(X - \bar{X})\), respectively. The matrix \(\Sigma \in {R}^{M \times N}\) is a diagonal matrix containing the singular values, which are non-negative real numbers arranged in descending order.

These singular values indicate the energy distribution in the data. Dimensionality reduction is achieved by truncating \(\Sigma\), retaining only the top \(k\) singular values and their corresponding vectors in \(U\) and \(V\), where \(k \ll N\).

The number of principal components \(k\) is determined by the cumulative energy criterion, defined as the ratio of the sum of the largest \(k\) eigenvalues to the total sum of all eigenvalues:

\begin{equation}\label{formula:CE}
    \text{Cumulative Energy} = \frac{\sum_{i=1}^{k} \lambda_i}{\sum_{i=1}^{N} \lambda_i}.
\end{equation}

\begin{figure}[htbp]
\centering
\begin{subfigure}[b]{0.6\linewidth}
\includegraphics[width=\linewidth]{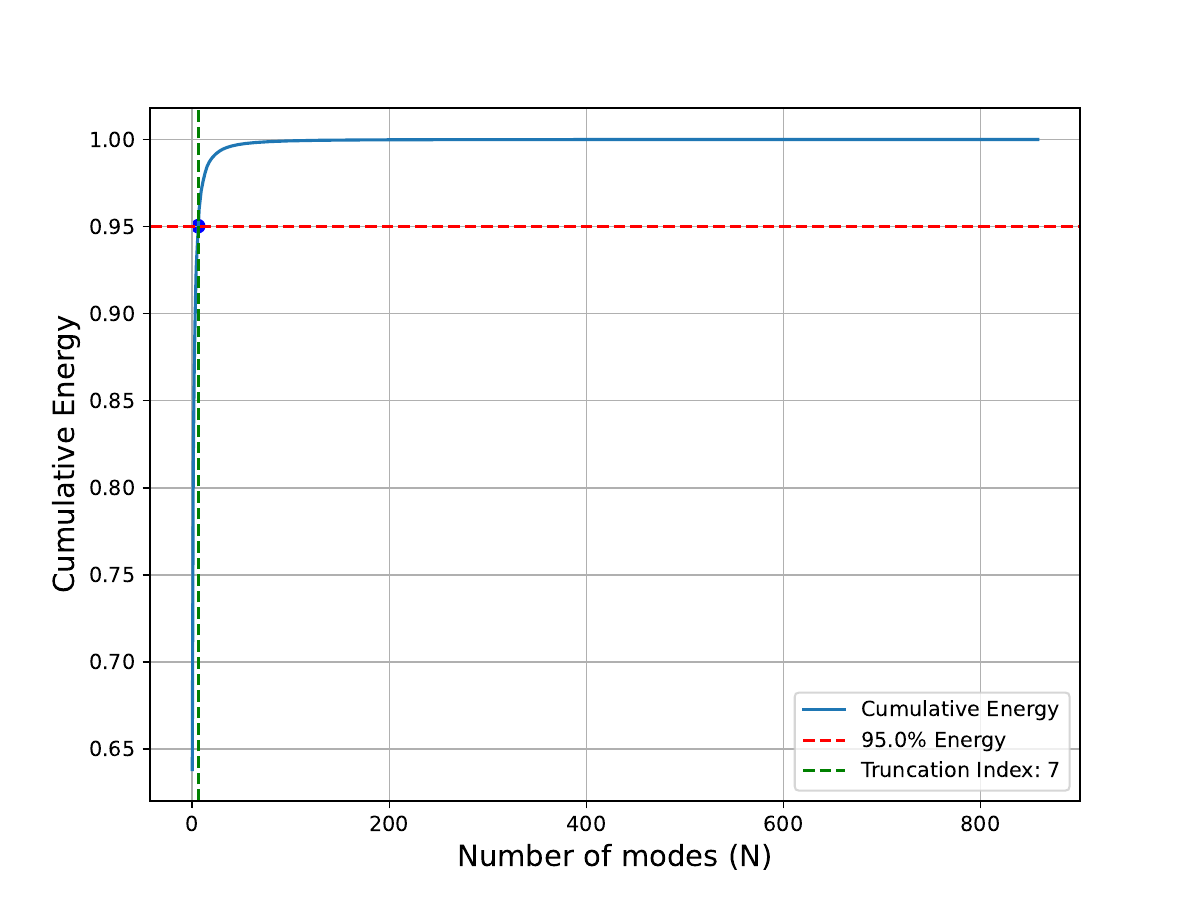}
\end{subfigure}
\caption{ The cumulative energy as a function of the number of modes, with the energy threshold set at 95\%, indicated by the red dashed line. The truncation index, \(k = 7\), is selected to capture the desired variance according to Eq. \ref{formula:CE}.
}
% run code /net/flood/data/users/malireza/bootstrap-crossvalid_GPU_version_asli.py
\label{fig:truncation}
\end{figure}

By setting an energy threshold of 95\%, we reduced the data's dimensionality to \(k = 7\), a significant decrease from the original \(N \approx 800\), as shown in Fig.~\ref{fig:truncation}. This step retains most of the variability in the data while greatly reducing computational complexity.

With PCA applied, the lower-dimensional matrices are then used to improve the efficiency of the EnsCGP model which can be approximated as:

\begin{equation}\label{Eq:GP}
     y_q \simeq \bar{Y} + \hat{Y} V_k \Sigma_k^{-1} U_k^{-1} (x_q - \bar{X}).
\end{equation}

The EnsCGP process effectively captures key regional attributes—such as river networks, topography, and other environmental characteristics—within its initial predictions. This inclusion of regional properties makes the EnsCGP-based initial estimate a robust candidate for initializing the ResNet.

\subsection{ResNet-18 }\label{sec:resnet}

We implemented the ResNet-18 model using input images with a resolution of $170 \times 130$ pixels, as shown in Fig. \ref{fig:area}
b. The model was developed using the PyTorch framework, and the training process was significantly expedited by utilizing NVIDIA A100 80GB PCIe GPUs.

The architecture of ResNet-18 consists of 18 layers organized into BasicBlock units, each comprising two convolutional layers. Each layer is followed by batch normalization and a Leaky ReLU activation function. To prevent overfitting during training, dropout regularization with a probability of 0.5 was applied. One of the key strengths of ResNet-18 is its identity shortcut connections, which allow gradients to flow more easily during backpropagation, mitigating the vanishing gradient issue commonly found in deeper networks.

The model starts with 64 filters in the first convolutional layer, and the number of filters doubles after each residual block, progressively increasing to 128, 256, and 512 filters. This gradual increase in filter size enables the model to capture spatial hierarchies from the input images effectively.

We used the RMSprop optimizer with a learning rate of \(5 \times 10^{-5}\) for training. The model was optimized using Smooth L1 Loss (also known as Huber Loss) as defined, which balances robustness to outliers with numerical stability. Smooth L1 Loss is defined as:

\begin{equation}
 \text{Smooth L1 Loss} = 
\begin{cases} 
0.5(\alpha - \beta)^2 & \text{if } |\alpha - \beta| < 1, \\
|\alpha - \beta| - 0.5 & \text{otherwise},
\end{cases}   
\end{equation}
where $\alpha$ displays the emulated water depths, while $\beta$ shows the corresponding flood simulation. The model was trained for 100 epochs with a batch size of 4, applying early stopping based on the validation set performance to prevent overfitting.

Following training, the ResNet-18 model was evaluated on the test dataset, processing each instance in approximately 0.001 seconds.

Since ResNet-18 was trained for 100 epochs and PCA was used to provide an initial guess in EnsCGP, the surrogate model training was efficient and could be quickly updated with new datasets.

\subsection{Bias Correction} \label{sec:bias}

Systematic biases in the surrogate model emulation are corrected using a quantile mapping method, applied on the empirical cumulative distribution function (ECDF) of flood depths. The biases are estimated by comparing the ECDF of simulated and emulated flood depths in the validation period, and the corrections are applied to the ECDF at the testing period. The corrected values of water depths are then back-projected from the corrected ECDF.

\section{Results}\label{sec:results}

To evaluate the surrogate model’s performance, we focused on two primary metrics: the infinity norm (\(L_{\infty}\)) and the coefficient of determination (\(R^2\)).

The infinity norm is expressed as:

\begin{equation}{\label{eq:err}}
\text{Error} = \max_{i} |y_i - \hat{y}_i|,
\end{equation}

where \(y_i\) is the flood simulation at pixel \(i\), and \(\hat{y}_i\) is the flood emulation from either the EnsCGP or ResNet-18 models, as shown in the workflow in Fig.~\ref{fig:flowchart}. This metric identifies the maximum pixel-wise error, offering insight into the worst-case performance of the model, which is particularly critical in flood depth prediction for assessing the model’s reliability over the entire spatial domain.

\begin{figure}[htbp]
\centering
\begin{subfigure}[b]{0.6\linewidth}
\includegraphics[width=\linewidth]{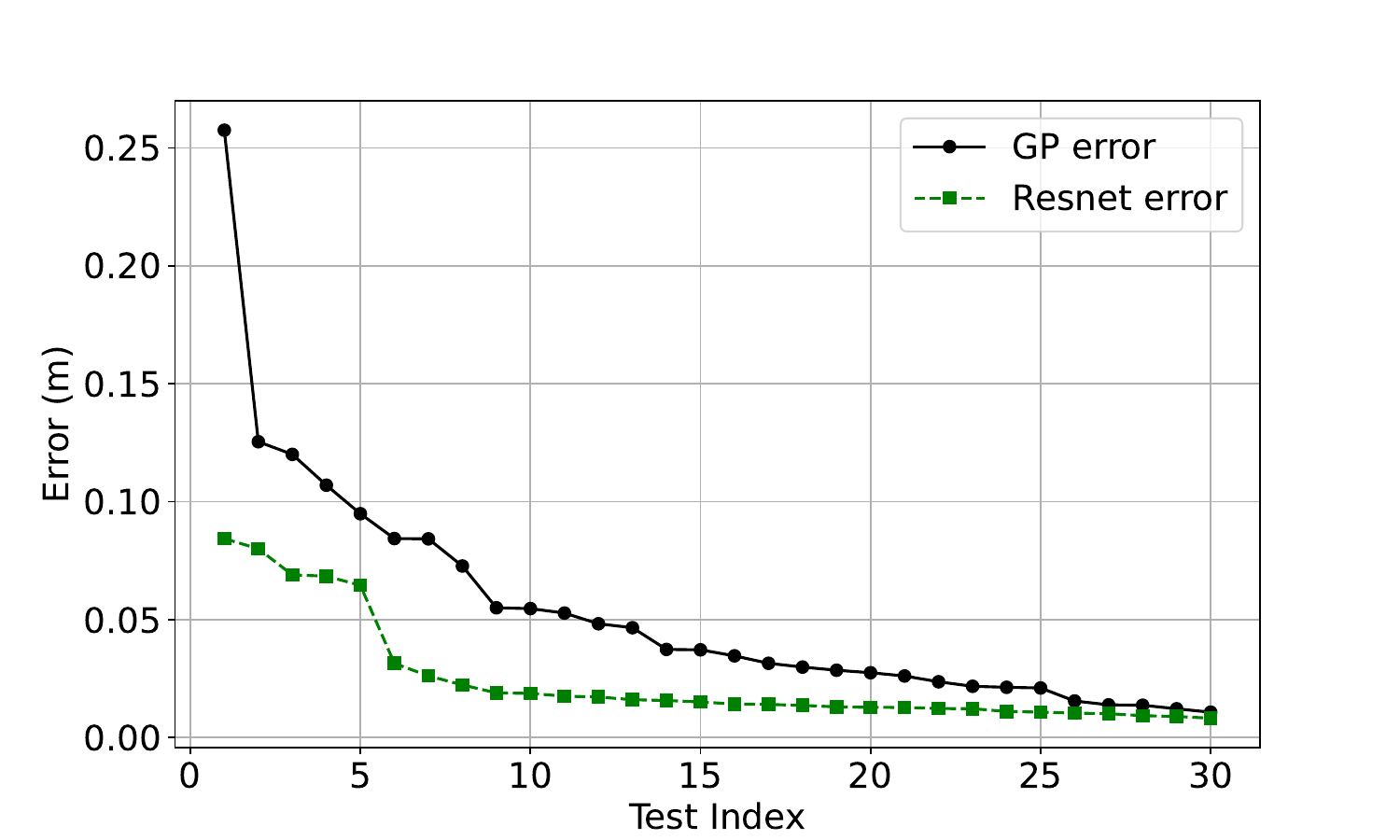}
\end{subfigure}
\caption{Maximum error values for the test set, indexed by event, comparing the performance of the Gaussian Process (EnsCGP) model (green) and the ResNet (black) for the year 2013. The plot highlights a notable reduction in error after applying the ResNet model, illustrating the neural network's ability to improve flood predictions initially produced by the EnsCGP model.
}
%rerun /net/flood/data/users/malireza/Bias_correction_ecdf.py
\label{fig:gp-res-err}
\end{figure}

Fig.~\ref{fig:gp-res-err} presents the maximum error comparison between EnsCGP and ResNet-18 for each test event in year 2013, highlighting the reduction in error achieved by ResNet-18. The black line represents the error calculated using Eq.~\ref{eq:err} for EnsCGP, while the green line corresponds to ResNet-18. No bias correction was applied to the ResNet-18 predictions in this comparison to emphasize the raw differences. The observed reduction in error illustrates how ResNet-18 improves upon EnsCGP by capturing finer spatial details, enhancing the accuracy of flood depth emulations.

In addition to the infinity norm, we used the coefficient of determination \(R^2\) to assess the overall predictive accuracy of the surrogate model on a pixel-by-pixel basis. The \(R^2\) value is calculated as:

\begin{equation}{\label{eq:crr}}
R^2 = 1 - \frac{\sum_{i=1}^{N} (y_i - \hat{y}_i)^2}{\sum_{i=1}^{N} (y_i - \bar{y})^2},
\end{equation}

where \(\bar{y}\) represents the mean of the flood simulation. The \(R^2\) metric quantifies how well the surrogate model's emulations align with the flood simulation, with values closer to 1 indicating stronger surrogate model performance.

The \(R^2\) score also reflects the proportion of variance in the flood simulation that the surrogate model can explain, providing a comprehensive measure of the surrogate model's predictive capability.

\subsection{Evaluation of Surrogate Model for year 2013} \label{sec:2013}

As an example, we consider the year 2013, an extreme flood year, to evaluate the surrogate model. However, in the current crop area (Figure \ref{fig:area}b), the maximum flood value is around 0.3 meters. As future work, we plan to extend this evaluation to crops containing extreme flood values of approximately 2.5 meters, as illustrated in Fig. \ref{fig:numerical_vs_gauge}b.

The dataset was divided into two main subsets: the training set, which spanned from February 10, 1981, to December 19, 2012, and a test set consisting of 30 specific flood events from 2013. The EnsCGP model was trained on the training set to generate the initial water depth predictions. To train ResNet-18, the training set was further split into a "network set" and a validation set, with 80\% of the data assigned to the network set for training the network, and 20\% reserved for validation. This distinction avoids any overlap in terminology with the training set used in the EnsCGP process.
The same splitting approach was applied in the section (\ref{sec:all year}).

\begin{figure}[htbp]
\centering
\begin{subfigure}[b]{0.4\linewidth}
\includegraphics[width=\linewidth]{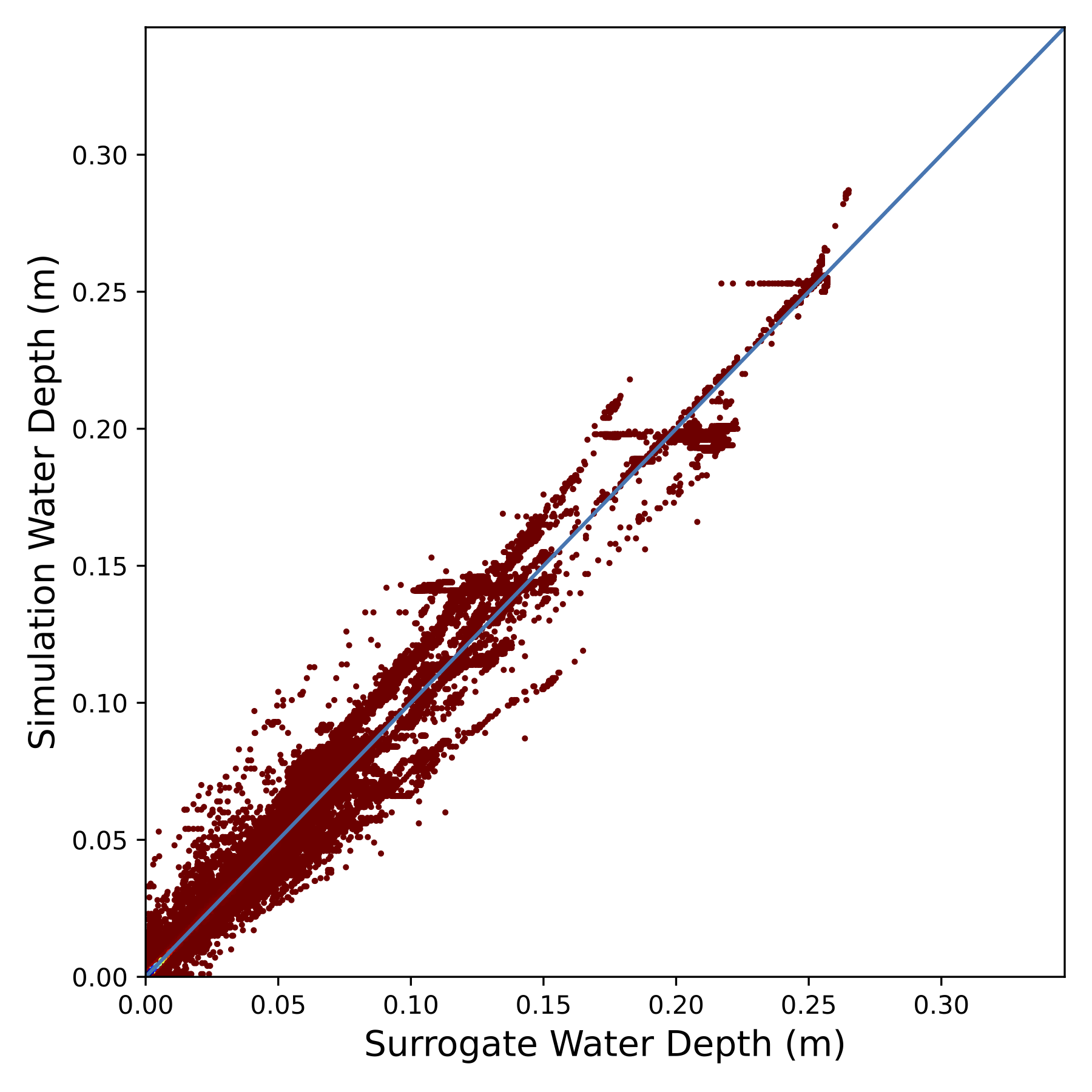}
\caption{Year 2013, \(R^2 = 0.98\)}
\end{subfigure}
\begin{subfigure}[b]{0.4\linewidth}
\includegraphics[width=\linewidth]%{fold_2013_val_test_max.pdf}
{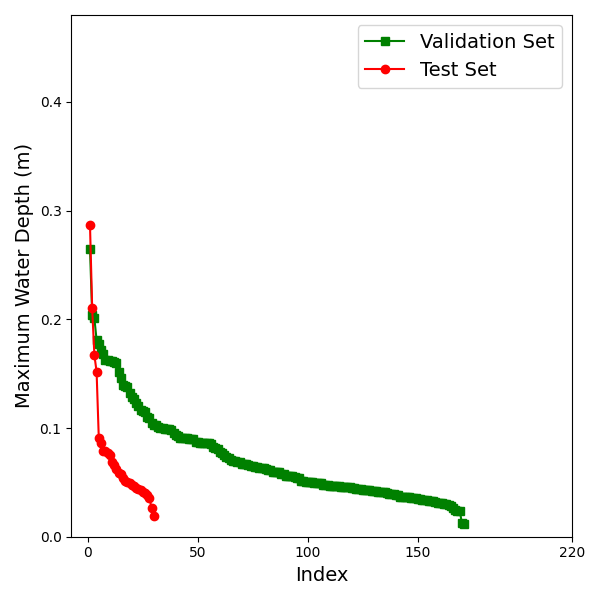}
\caption{water depth}
\end{subfigure}
% code for both is at /net/flood/data/users/malireza/Bias_correction_ecdf.py and the data is at /net/flood/data/users/malireza/crossvalid_surrogate_model_Resnet18_valid_asli.py
\caption{(a) Scatter plot comparing simulated water depths with those predicted by the surrogate model for 2013. The data points align closely with the diagonal, indicating strong agreement between the predictions and simulations. (b) Line plot showing maximum water depths across validation and test sets. Despite the test set displaying higher maximum depths, the surrogate model accurately predicts these values, demonstrating its robustness in handling extreme conditions.}

\label{fig:2013scatter-num-pred}
\end{figure}

\begin{figure}[htbp]
\centering
\begin{subfigure}[b]{0.32\linewidth}
% \caption{Surrogate}
% \includegraphics[width=\linewidth]{Y_prd_2013.png}
\includegraphics[width=\linewidth]{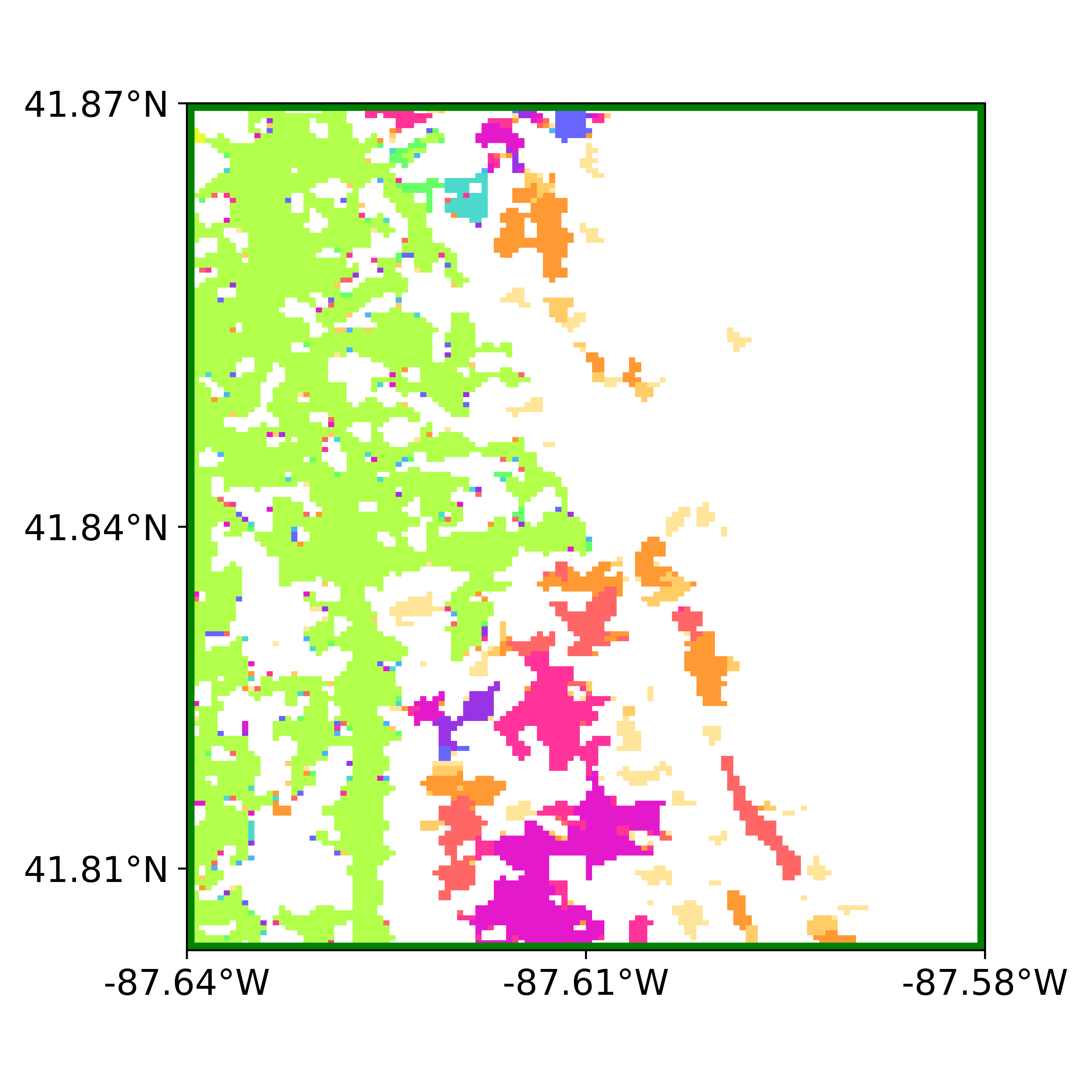}
\end{subfigure}
\begin{subfigure}[b]{0.32\linewidth}
% \caption{Simulation}
% \includegraphics[width=\linewidth]{Y_tst_2013.png}
\includegraphics[width=\linewidth]{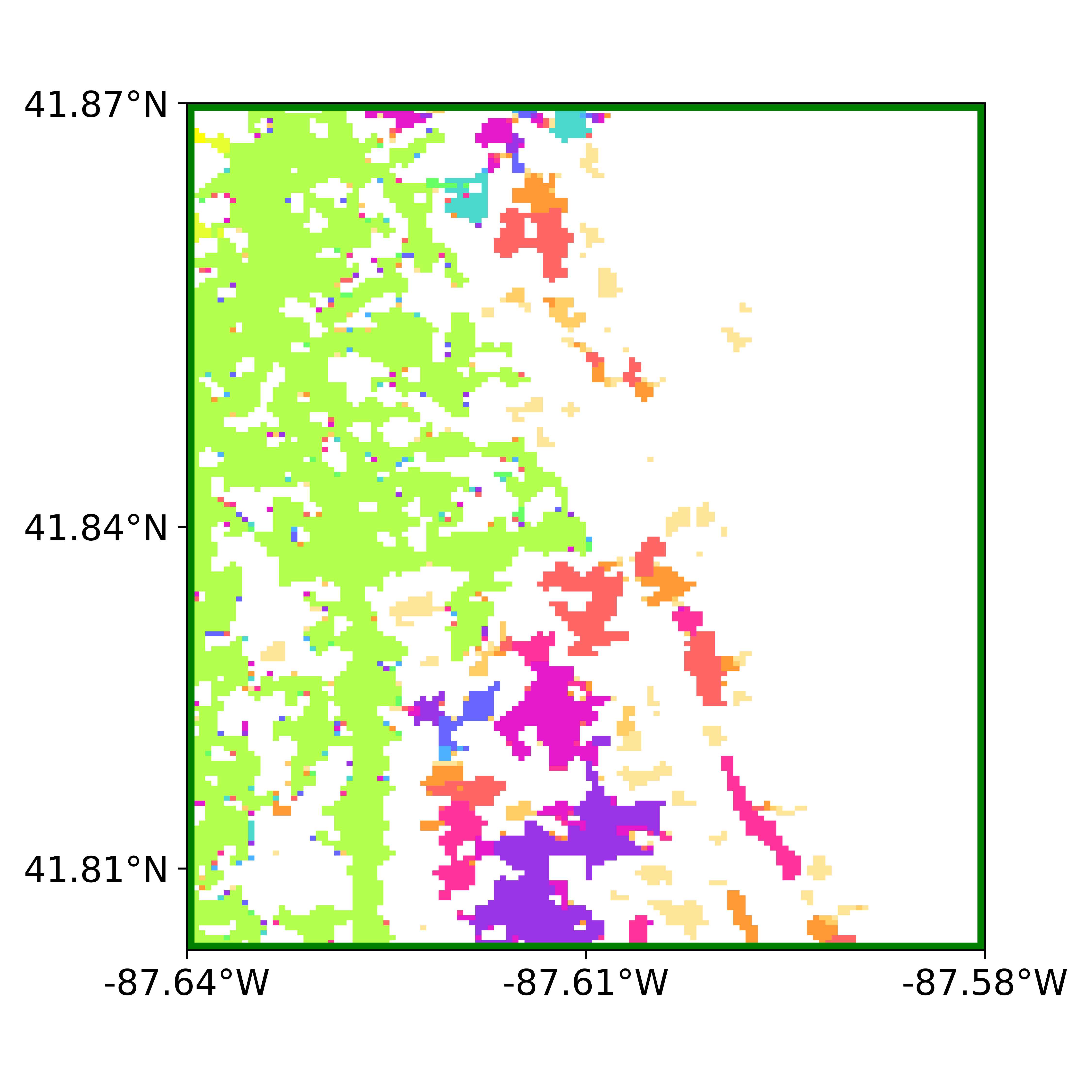}
\end{subfigure}
\begin{subfigure}[b]{0.32\linewidth}
% \caption{Difference}
% \includegraphics[width=\linewidth]{Y_diff_2013.png}
\includegraphics[width=\linewidth]{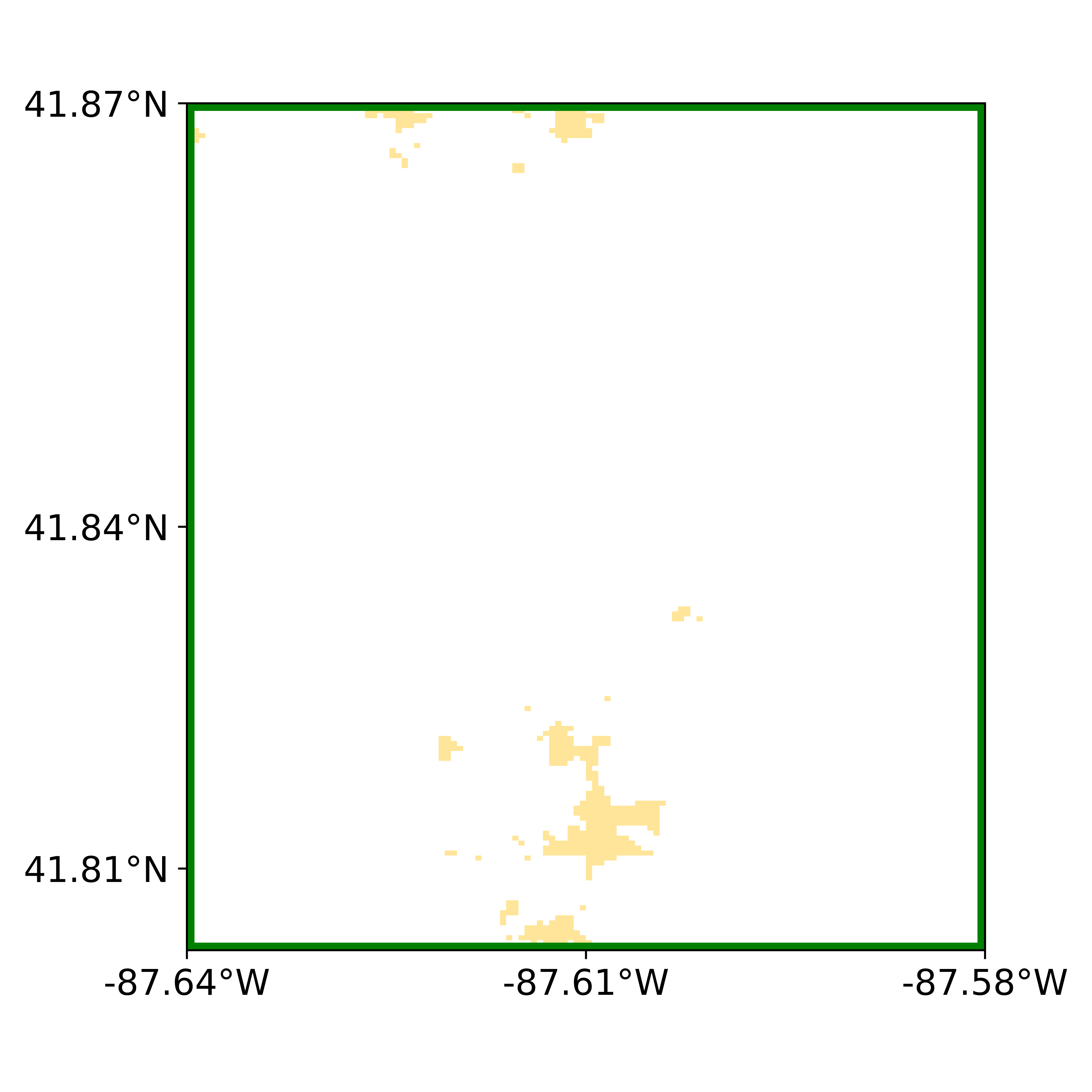}
\end{subfigure}
% Second row: Horizontal colorbar
\includegraphics[width=0.5\linewidth]{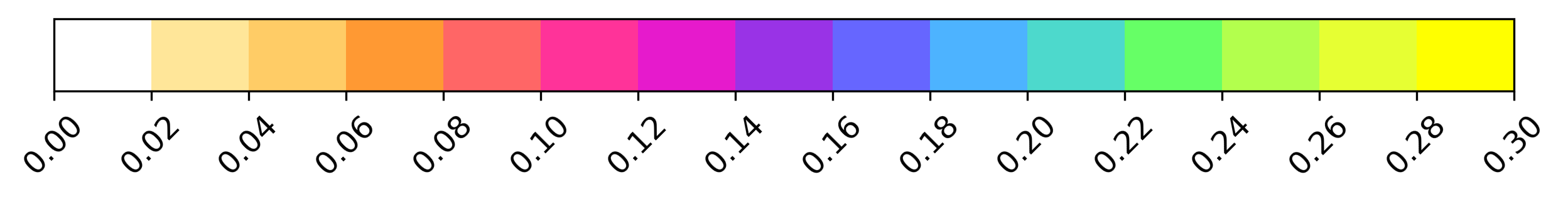}

%code map is in /net/flood/data/users/malireza/image_map.py and data is from /net/flood/data/users/malireza/Bias_correction_ecdf.py
\caption{Maximum flood inundation (mater) maps for downtown Chicago in 2013, comparing the flood emulations (left) with the flood simulation results (middle) and the absolute value difference between Surrogate and numerical (right). }
\label{fig:flood map depth}
\end{figure}

Despite the test data containing water depths beyond the range used for validation, the model was able to predict the flood depths effectively. In Fig.~\ref{fig:2013scatter-num-pred}(a), the scatter plot compares water depths emulated by the surrogate model with the simulation results. The high \(R^2\) value of 0.98 shows a strong match between the two, proving the model’s ability to predict flood depths accurately. Fig.~\ref{fig:2013scatter-num-pred}(b) shows the maximum water depths from both the test and validation sets. Although some test data points exceed the training range, the model captures these extreme cases well, demonstrating that it can generalize beyond the training data. Finally, Fig.~\ref{fig:flood map depth} shows flood maps for downtown Chicago (Fig.~\ref{fig:area} b) during the worst flood event of 2013. The comparison between the surrogate model emulations (Fig.\ref{fig:flood map depth}
a) and the numerical simulations (Fig.\ref{fig:flood map depth}
b) shows minimal differences, as highlighted in the residual map (Fig.~\ref{fig:flood map depth}c), demonstrating the surrogate model's strong alignment with the numerical simulations. This figure also shows the locations and extent of the maximum water depth, and in the difference map, we can observe where the surrogate model performs well in emulation and where it falls short.

According to Fig.~\ref{fig:numerical_vs_gauge}b, when we compared the observation versus simulation for the region (Fig.~\ref{fig:area}a) and then compared the simulation with emulation for the region (Fig.~\ref{fig:area}b) in the scatter plot (Fig.~\ref{fig:2013scatter-num-pred}) with $R^2=0.98$, we can conclude how surrogate results can be comparable with the observation.

% \begin{figure}[htbp]
% \centering
% \begin{subfigure}[b]{0.4\linewidth}
% \includegraphics[width=\linewidth]{real_num_scatter2013.png}
% \end{subfigure}
% \caption{Scatter plot comparing the observation with the simulation water depth for the year 2013.}
% \label{fig:2013scatter-num-real}
% \end{figure}

\subsection{Evaluation of Surrogate Model Across Multiple Years}\label{sec:all year}

The surrogate model's performance was evaluated by comparing its emulations with numerical simulations over several years, aiming to validate the model's accuracy, generalization capability, and computational efficiency across varying flood conditions.

\begin{figure}[htbp]
\centering
\begin{subfigure}[b]{0.4\linewidth}
\includegraphics[width=\linewidth]{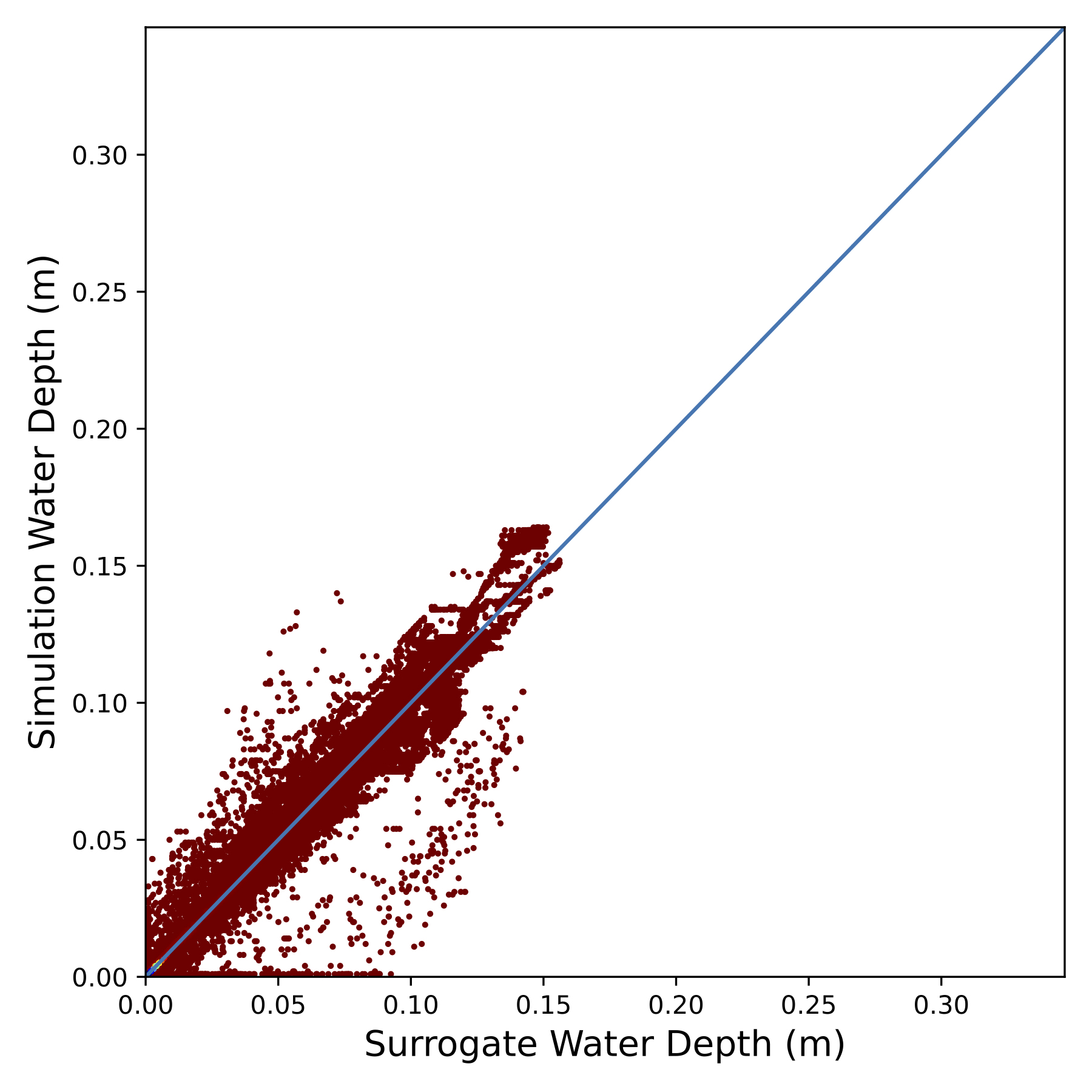}
\caption*{Year 2014, $R^2=0.96$}
\end{subfigure}
\begin{subfigure}[b]{0.4\linewidth}
\includegraphics[width=\linewidth]{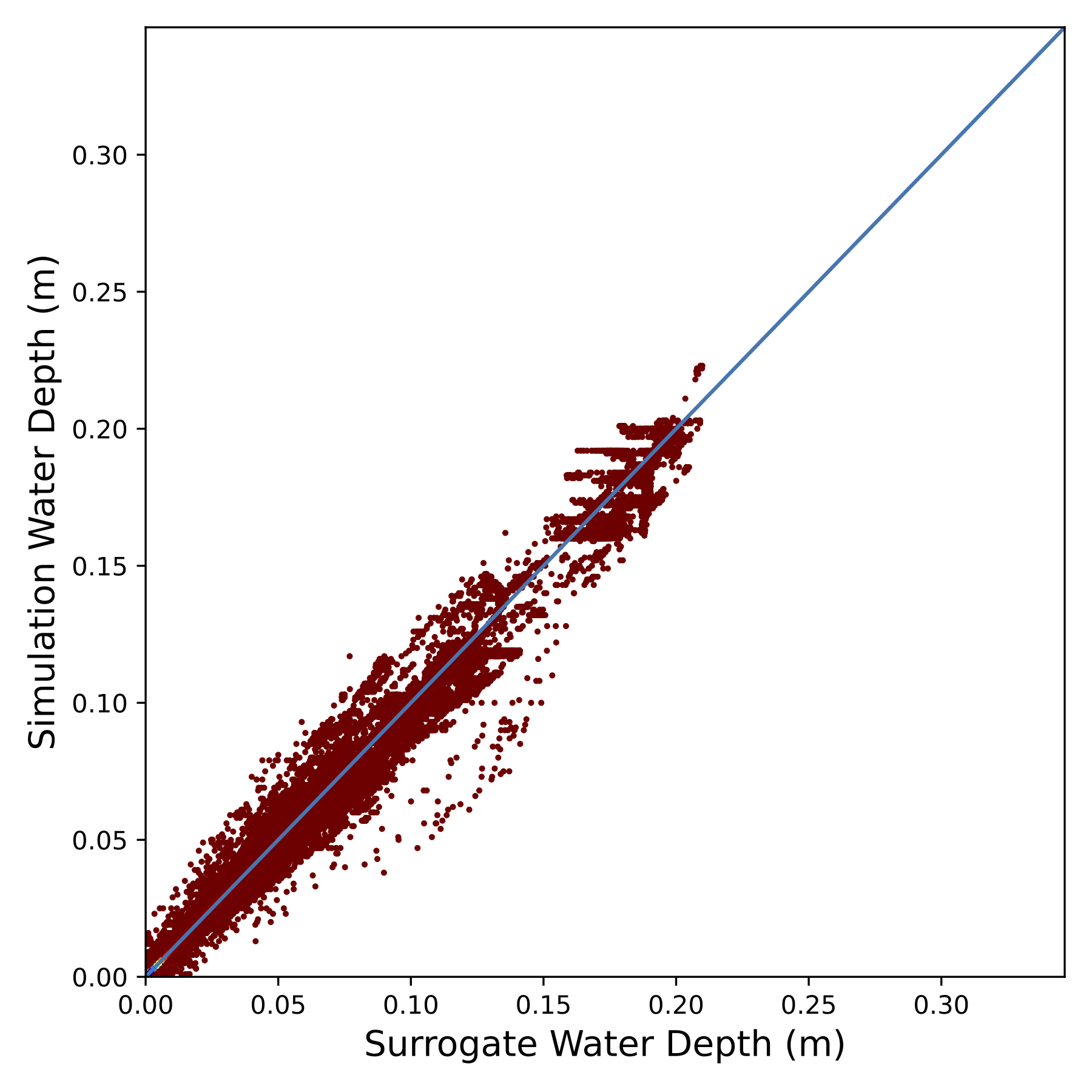}
\caption*{Year 2015, $R^2= 0.99$}
\end{subfigure}
\begin{subfigure}[b]{0.4\linewidth}
\includegraphics[width=\linewidth]{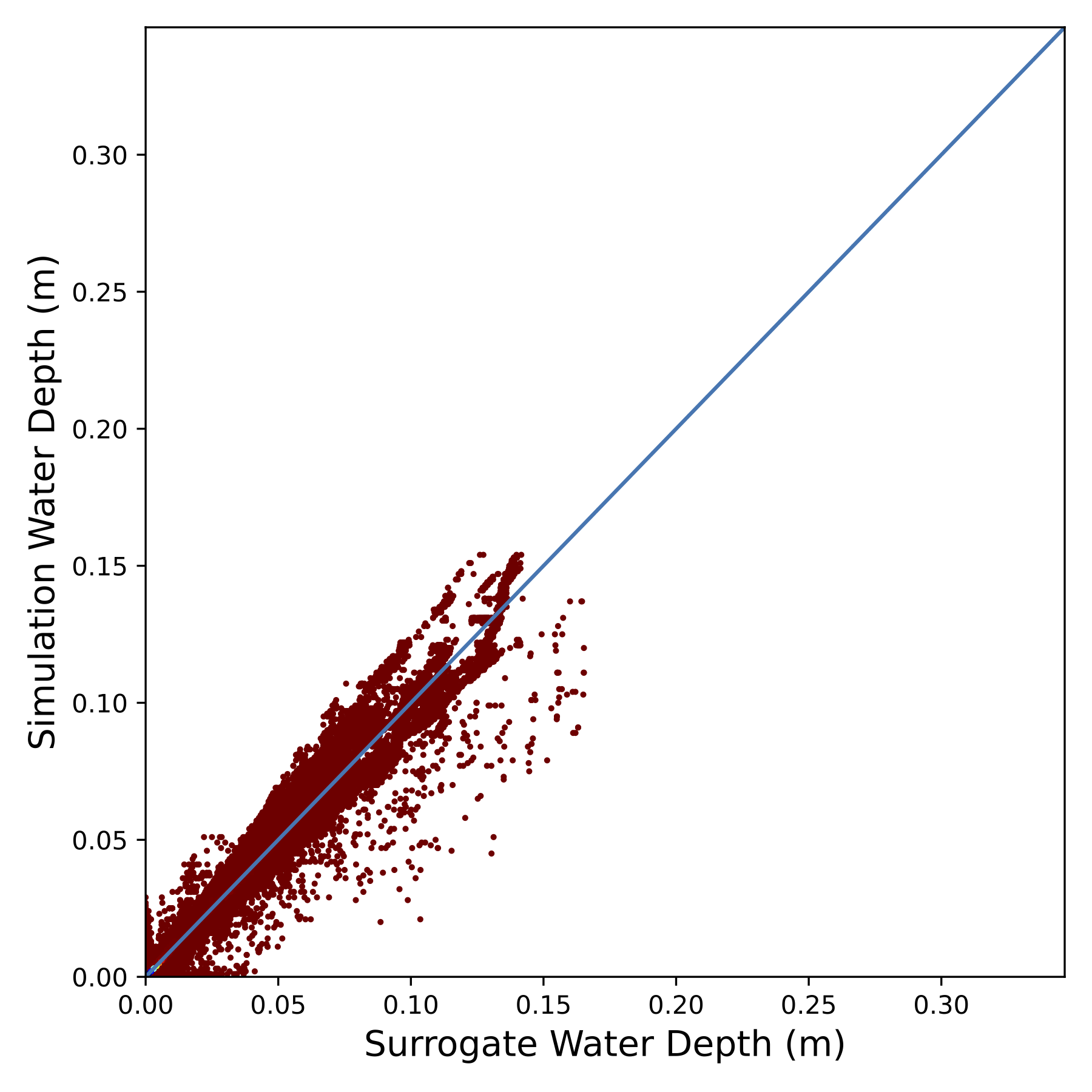}
\caption*{Year 2016, $R^2= 0.97$}
\end{subfigure}
\begin{subfigure}[b]{0.4\linewidth}
\includegraphics[width=\linewidth]{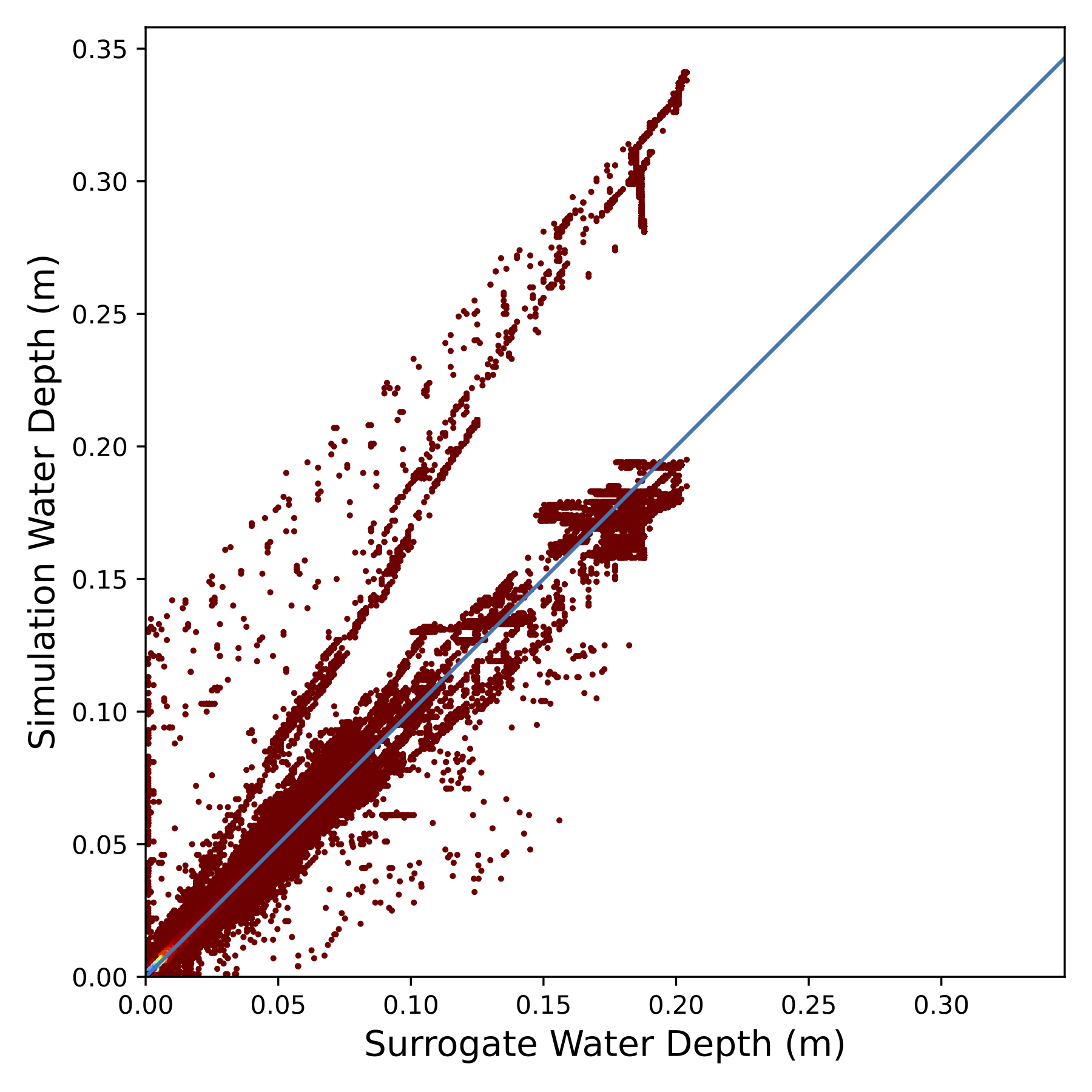}
\caption*{Year 2017, $R^2=0.89$}
\end{subfigure}
\begin{subfigure}[b]{0.4\linewidth}
\includegraphics[width=\linewidth]{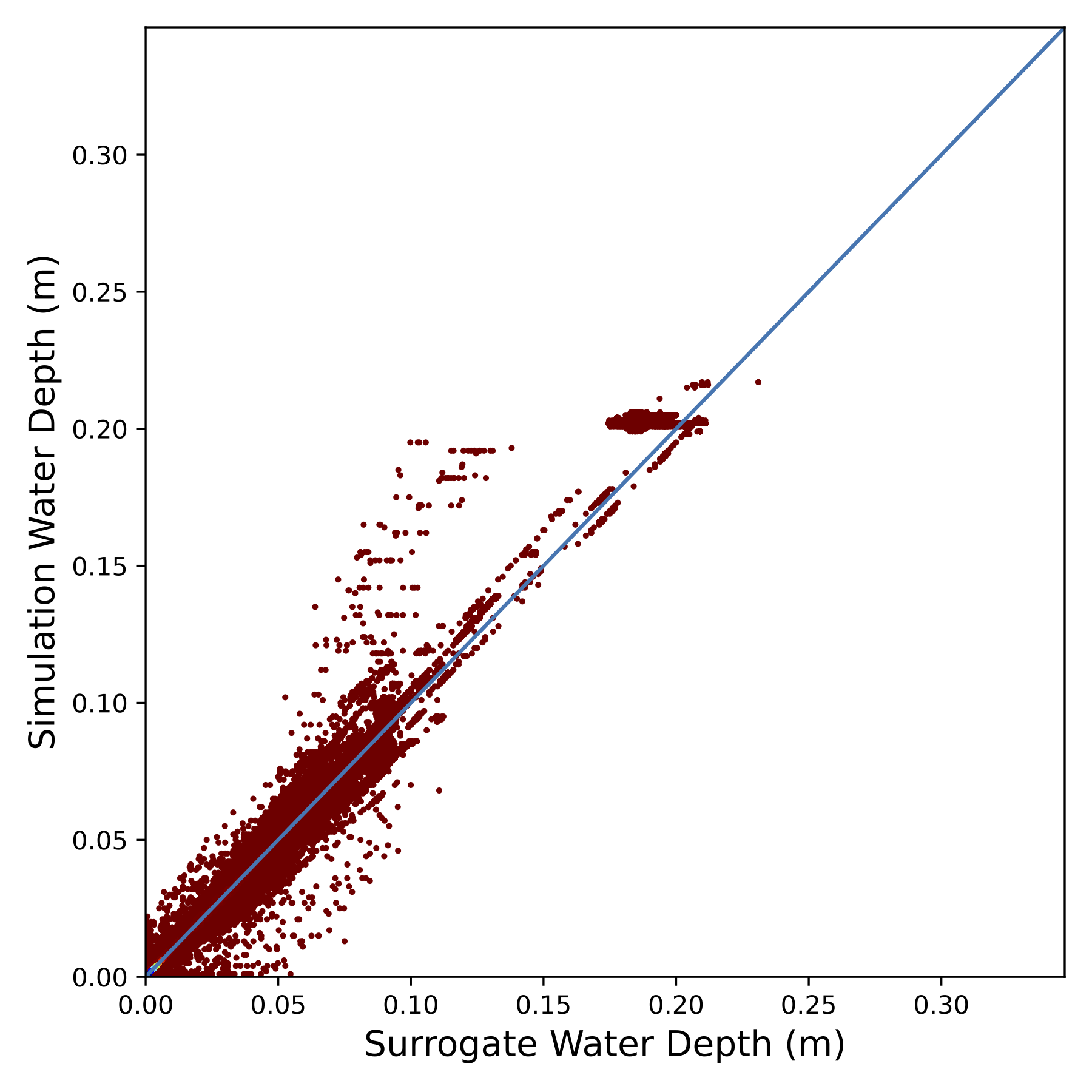}
\caption*{Year 2018, $R^2=0.98$}
\end{subfigure}
\begin{subfigure}[b]{0.4\linewidth}
\includegraphics[width=\linewidth]{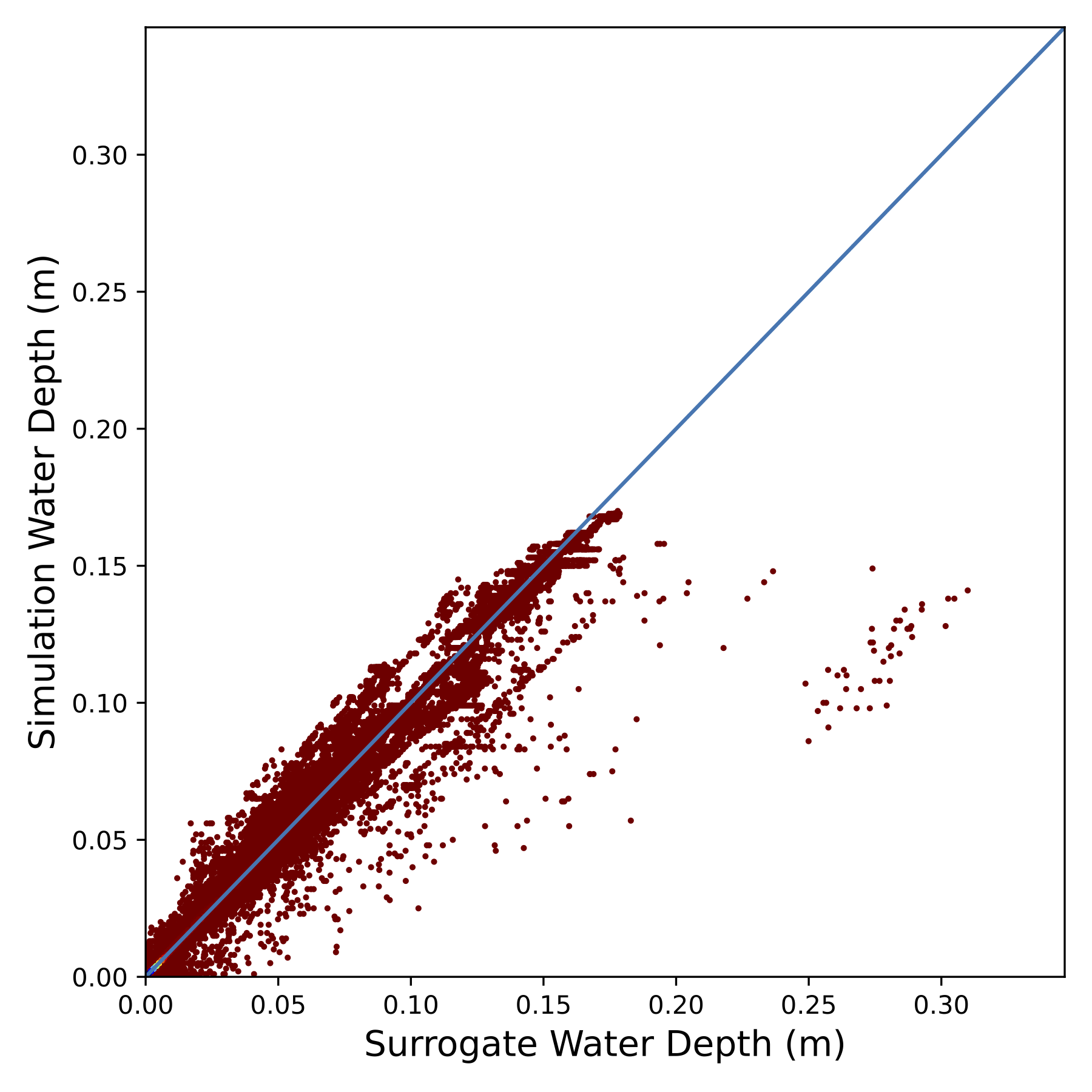}
\caption*{Year 2019, $R^2=0.98$}
\end{subfigure}
\caption{Scatter plots for 2014 to 2019 illustrate the relationship between simulated flood depths and surrogate model emulations . The $R^2$
  values demonstrate consistently high correlations across all years. Particularly in 2017, even when the maximum flood depths in the test set exceeded those of the validation set (as seen in Fig. \ref{fig:valid_test}), the model maintained its capacity to predict accurate flood depths.
 }
\label{fig:scatter-allyears}
\end{figure}

Fig. \ref{fig:scatter-allyears} presents scatter plots comparing the surrogate model emulations to flood simulation for 2014 to 2019. The results show a strong correlation, with the coefficients of determination $R^2$
 demonstrating consistent alignment between the two approaches. This high correlation across multiple years demonstrates the robustness of the surrogate model.

\begin{figure}[htbp]
\centering
\begin{subfigure}[b]{0.4\linewidth}
\includegraphics[width=\linewidth]%{fold_2014_val_test_max.pdf}
{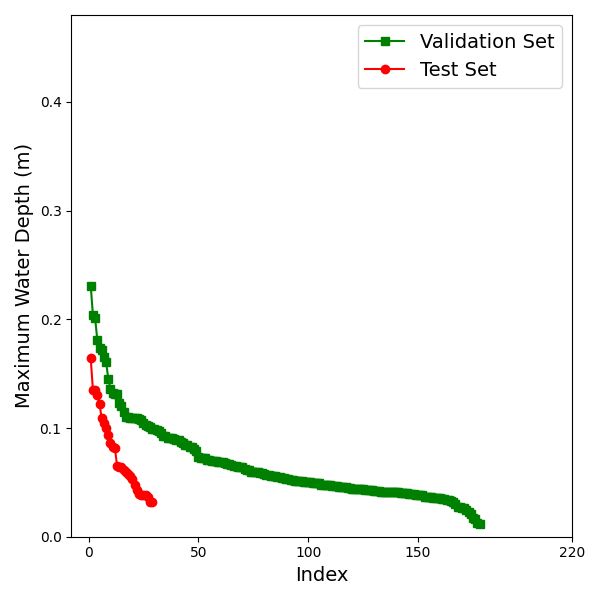}
\caption*{Year 2014}
\end{subfigure}
\begin{subfigure}[b]{0.4\linewidth}
\includegraphics[width=\linewidth]%{fold_2015_val_test_max.pdf}
{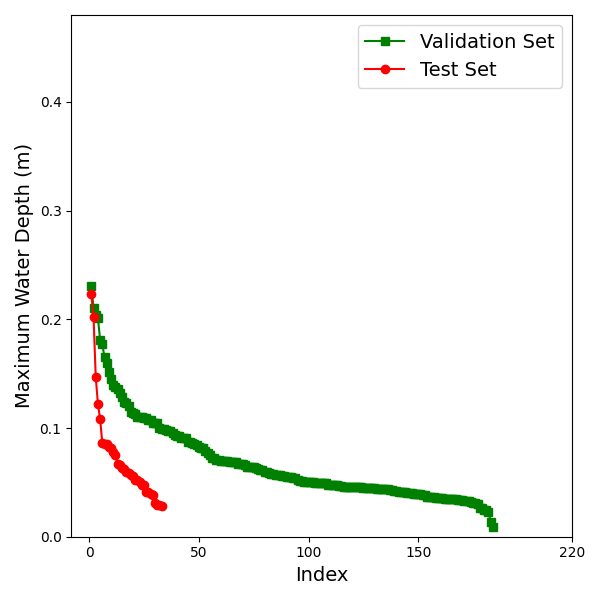}
\caption*{Year 2015}
\end{subfigure}
\begin{subfigure}[b]{0.4\linewidth}
\includegraphics[width=\linewidth]%{fold_2016_val_test_max.pdf}
{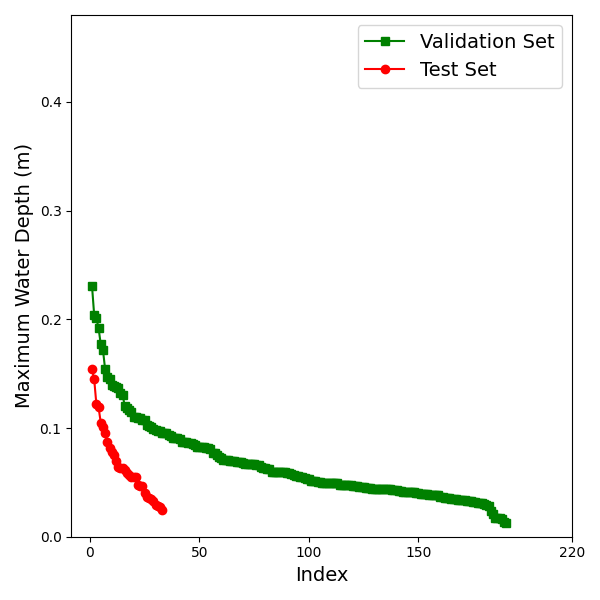}
\caption*{Year 2016}
\end{subfigure}
\begin{subfigure}[b]{0.4\linewidth}
\includegraphics[width=\linewidth]%{fold_2017_val_test_max.pdf}
{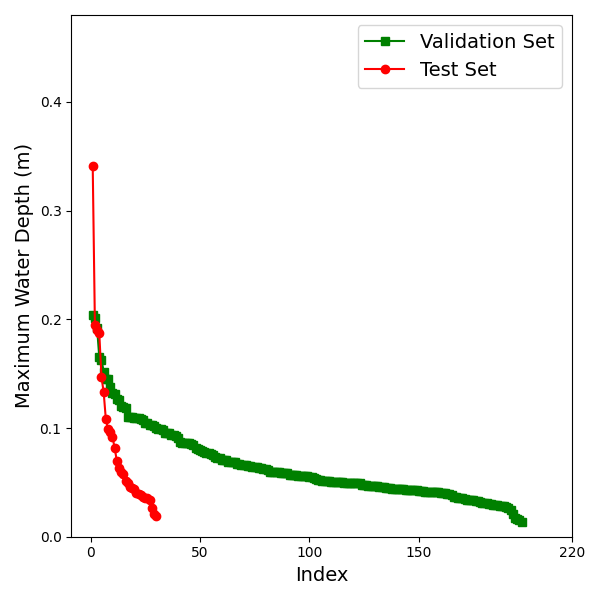}
\caption*{Year 2017}
\end{subfigure}
\begin{subfigure}[b]{0.4\linewidth}
\includegraphics[width=\linewidth]%{fold_2018_val_test_max.pdf}
{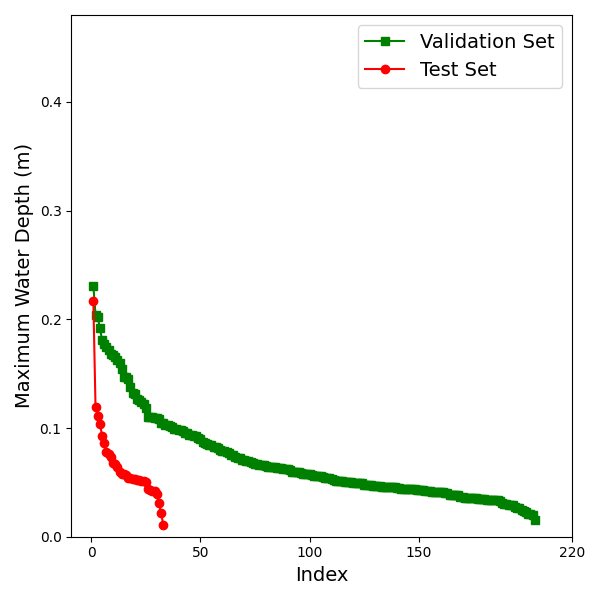}
\caption*{Year 2018}
\end{subfigure}
\begin{subfigure}[b]{0.4\linewidth}
 \includegraphics[width=\linewidth]%{fold_2019_val_test_max.pdf}
{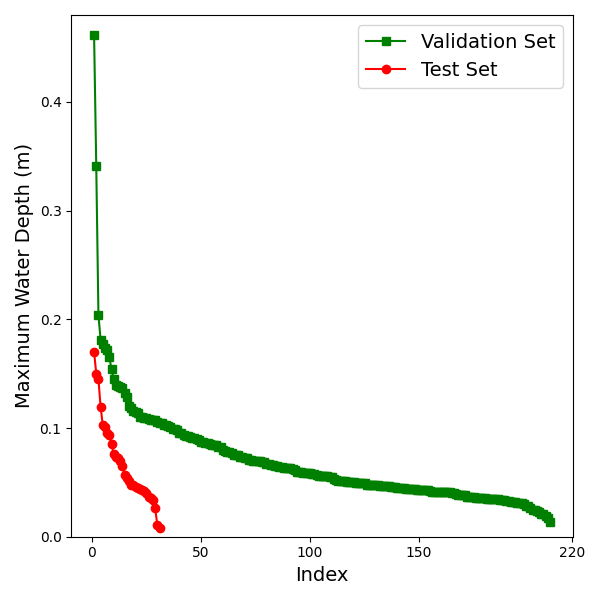}
\caption*{Year 2019}
\end{subfigure}
    \caption{Maximum flood inundation values (meter) are shown over validation set (green) and  test set (red)  indices for the years 2014 to 2019. In 2017, the test set exhibits some large flood events that were not present in the validation data. } 
    \label{fig:valid_test}
    % run code /net/flood/data/users/malireza/Bias_correction_ecdf.py
\end{figure}

The surrogate model demonstrates optimal performance when the maximum water depth in the test set is similar to or lower than that of the validation set (Fig. \ref{fig:valid_test}
), as indicated by well-aligned scatter plots and $R^2$ values close to one. However, when the test set contains higher water depths than the validation set —for instance, in year 2017, where the maximum test water depth reached 0.35 meters, exceeding the validation set’s maximum of 0.15 meter—the scatter plot deviates from the ideal 45-degree line, reflecting a slight decrease in prediction accuracy.

\begin{figure}[htbp]
    \centering

        % First Row
        \includegraphics[width=\linewidth]{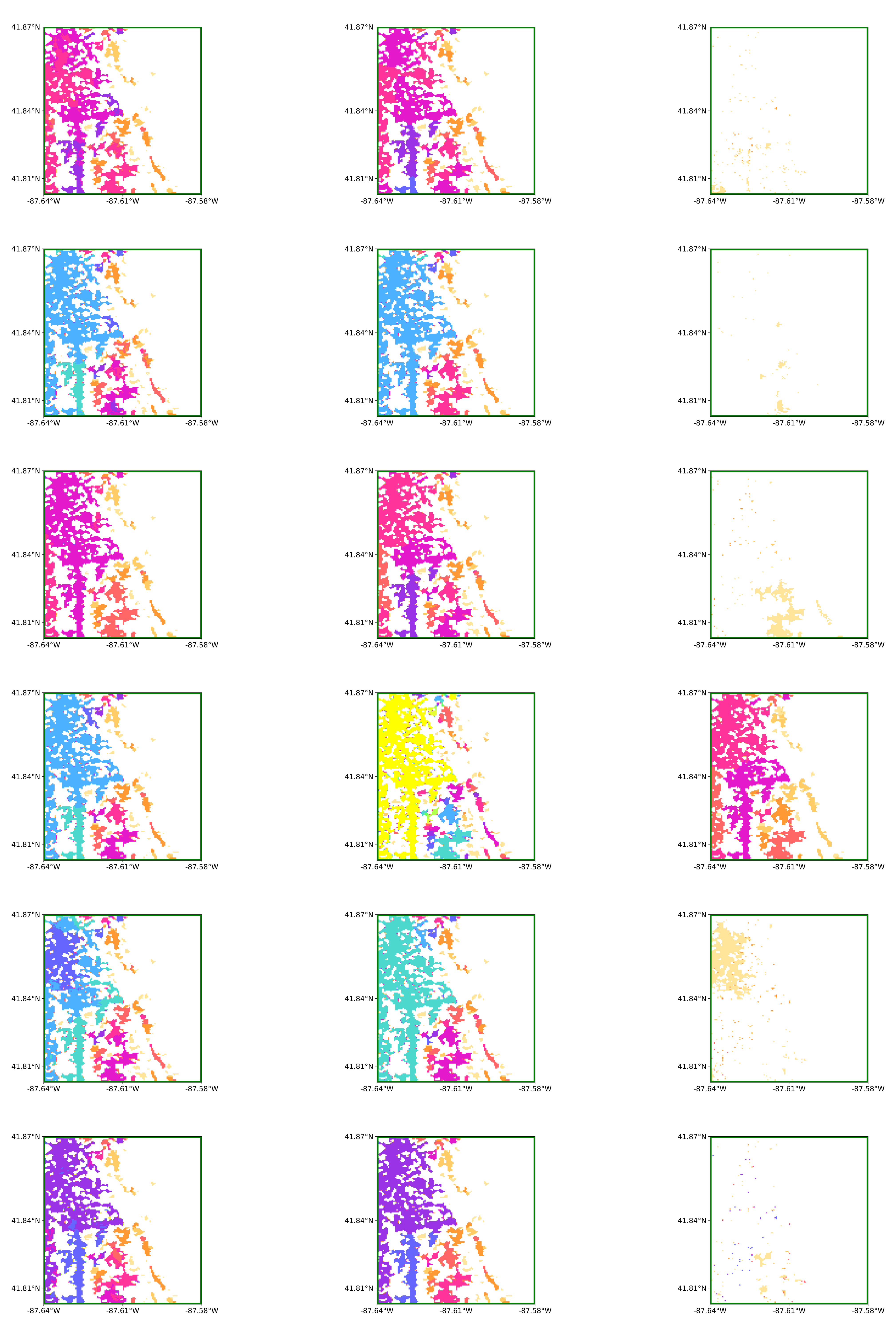} \\
        % \includegraphics[width=0.9\linewidth]{2015.png} \\
        % \includegraphics[width=0.9\linewidth]{2016.png} \\
        % % % Second Row
        % \includegraphics[width=0.9\linewidth]{2017.png} \\
        % \includegraphics[width=0.9\linewidth]{2018.png} \\
        % \includegraphics[width=0.9\linewidth]{2019.png} 
        \includegraphics[width=0.5\linewidth]{colorbar.png}
        \caption{Maximum flood inundation maps for downtown Chicago arranged in a vertical sequence from 2014 (top) to 2019 (bottom), comparing the highest water depth for each year. The comparisons are shown between the flood emulations (left), flood simulation (middle), and the absolute differences between them (right).}
\label{fig:all flood map depth}
 \end{figure}

Fig.~\ref{fig:all flood map depth} provides a detailed comparison of the absolute differences between the surrogate emulations and the numerical simulation for the maximum flood event in each year from 2014 to 2019. The first column presents the flood inundation maps generated by the surrogate model, the second column shows the corresponding numerical simulation results, and the third column illustrates the pixel-wise absolute differences between the two. 

\begin{figure}[htbp]
\centering
\begin{subfigure}[b]{0.8\linewidth}
\includegraphics[width=\linewidth]{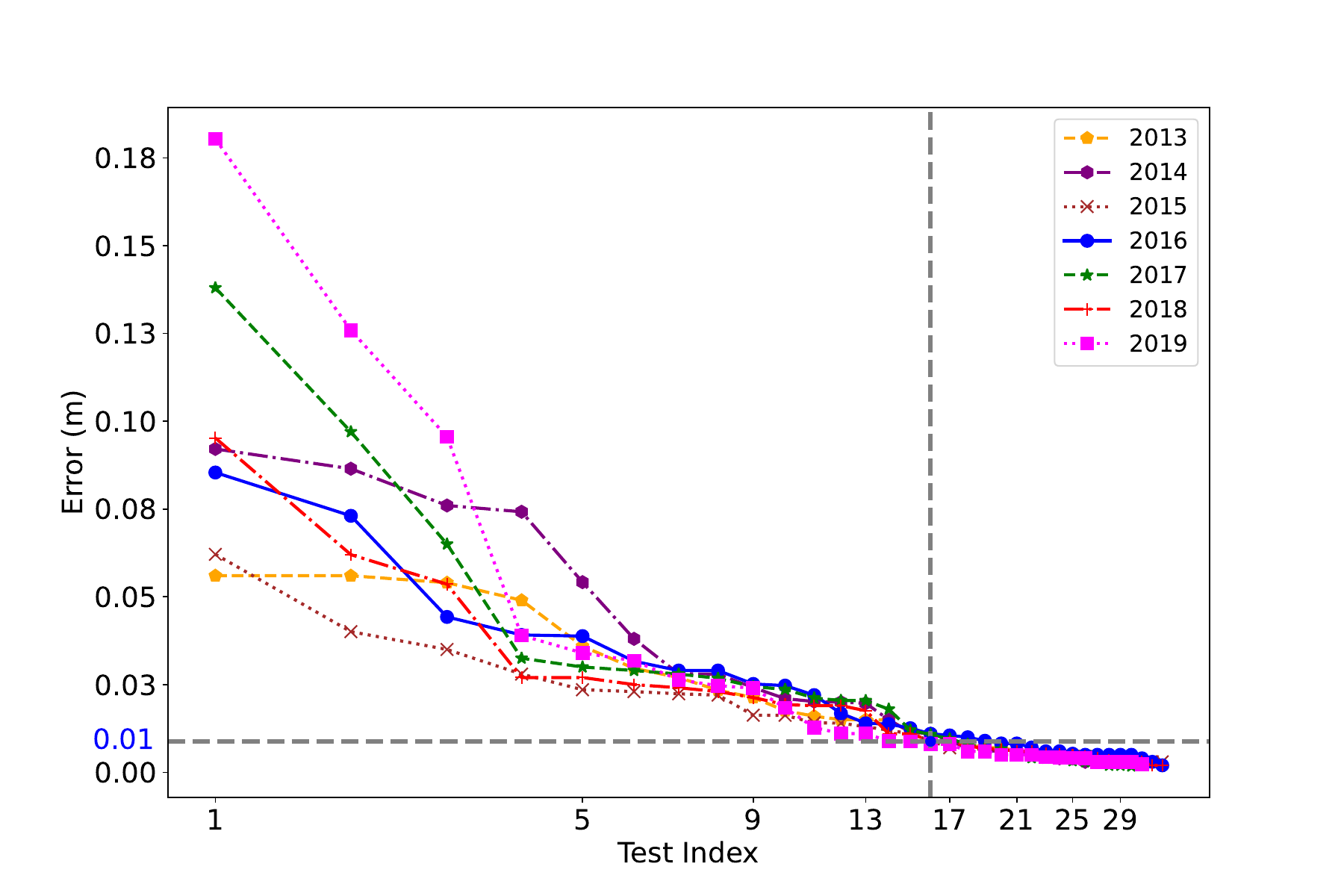}
\end{subfigure}
\caption{Error plots comparing surrogate model predictions with numerical simulations for 2013-2019. The plots show error convergence to low values, with median flood depth errors around 1 cm, highlighting the model's robustness.}
\label{fig:error-allyears}
\end{figure}

In Fig.~\ref{fig:error-allyears}, we plot the maximum error for all dates from 2013 to 2019 in a single plot, allowing us to assess the median error of the surrogate model over multiple years. The error plots show that the median water depth error across all years is approximately 1 cm, demonstrating the model’s consistency and reliability in emulating flood inundation.

The combination of high \(R^2\) values and minimal error convergence across various conditions confirms the surrogate model’s effectiveness in providing accurate and computationally efficient predictions. These results showcase the model's potential for practical applications in flood risk management, where timely and accurate predictions are critical for mitigating flood-related risks.

%\subsection{Optional: Cross‐Validation}
%We propose adding a section on bootstrapping to evaluate the ResNet model's performance. Specifically, we will create 20 different random selections of the ResNet training set for each year from 2013 to 2019. In this approach, the GP component remains fixed (frozen) for each year, and only the ResNet model is retrained using the different random splits. This allows us to isolate and assess the ResNet's performance independently by varying its training data.
\section{Conclusions}{\label{sec:conclusion}}

This study introduced a surrogate model integrating ensemble-approximated Conditional Gaussian Processes (EnsCGP) with a ResNet-18 deep learning network to emulate flood inundation across varied scenarios efficiently. Trained on historical rainfall data from the Daymet dataset and flood simulations produced by LISFLOOD-FP, the surrogate model demonstrated robust performance in emulating  flood events. This approach significantly reduced computational costs compared to traditional hydrodynamic models while maintaining high accuracy. Furthermore, bias correction enhanced model precision, ensuring reliable generalization to unseen flood conditions.

% Future research will focus on extending the applicability of this surrogate model beyond inland pluvial flooding. While primarily applied to this type of flooding, the model's adaptability allows for potential extensions to coastal areas using models like SFINCS \cite{LEIJNSE2021103796} and applicability to regions worldwide using datasets such as CHIRPS \cite{Funk_2015}. Additionally, the model is designed to accommodate current and future climate scenarios, including CMIP6 and HighResMIP  projections \cite{eyring2016overview}, offering a versatile tool for flood damage assessment, loss estimation, and proactive disaster management. Its demonstrated utility in assessing inland flood risk underscores its relevance for insurance applications and the broader goal of mitigating the impacts of a changing climate.

Future research will focus on extending the applicability of this surrogate model to other regions, including areas with maximum flood depths around 3 meters. By applying the model to these larger flood scenarios, we aim to demonstrate its versatility and robustness in handling more extreme flood events. While the current study primarily addressed inland pluvial flooding, the model's adaptability allows for potential extensions to coastal areas using models like SFINCS \cite{LEIJNSE2021103796} and global applications using datasets such as CHIRPS \cite{Funk_2015}.

Additionally, the model is designed to accommodate current and future climate scenarios, including CMIP6 and HighResMIP projections \cite{eyring2016overview}. This adaptability offers a versatile tool for flood damage assessment, loss estimation, and proactive disaster management. The demonstrated utility in assessing inland flood risk underscores its relevance for insurance applications and contributes to the broader goal of mitigating the impacts of a changing climate. 

\section{Acknowledgments}

The authors gratefully acknowledge the support provided by Liberty Mutual (029024-00020), ONR (N00014-19-1-2273).

% \section{CRediT authorship contribution statement}
% \textbf{Marzieh Alireza Mirhoseini}: Software, Validation, Formal analysis, Investigation, Data curation, Writing $-$  original draft, Visualization. \textbf{Anamitra Saha}: Software, Validation, Writing $-$ review and editing. \textbf{Sai Ravela}: Conceptualization, Methodology, Resources, Writing $-$ review and editing, Supervision, Project administration, Funding acquisition.

\bibliographystyle{unsrt}
\bibliography{flood}

\end{document}